\RequirePackage{fix-cm}
\documentclass[smallextended]{svjour3}       
\smartqed  
\usepackage{graphicx}
\usepackage{natbib} 
\usepackage[colorlinks=true, linkcolor=blue, citecolor=blue, urlcolor=blue]{hyperref}
\usepackage{amsmath} 
\usepackage{booktabs} 
\usepackage{pifont}
\newcommand{\cmark}{\ding{51}} 
\newcommand{\xmark}{\ding{55}} 
\usepackage[most]{tcolorbox} 
\usepackage{rotating} 

\usepackage{orcidlink}
\newcommand{\orcid}[1]{\,\textcolor[HTML]{A6CE39}{\orcidlink{#1}}}

\usepackage{silence}
\usepackage{subfig}

\usepackage{microtype}  
\journalname{Empirical Software Engineering}
\begin{document}

\title{An Empirical Study of the Realism of Mutants in Deep Learning}

\author{Zaheed Ahmed\orcid{0000-0001-6594-537X}
        \and
        Philip Makedonski\orcid{0000-0001-7752-0029}
        \and
        Jens Grabowski\orcid{0000-0003-2994-3531} 
}


\institute{Z. Ahmed \at
              Institute of Computer Science, University of Goettingen, Goettingen, Germany \\
              \email{zaheed.ahmed@informatik.uni-goettingen.de}           
           \and
           P. Makedonski \at
              Institute of Computer Science, University of Goettingen, Goettingen, Germany \\
              \email{makedonski@informatik.uni-goettingen.de}
           \and
           J. Grabowski \at
              Institute of Computer Science, University of Goettingen, Goettingen, Germany \\
              \email{grabowski@informatik.uni-goettingen.de}
}

\date{Received: date / Accepted: date}

\maketitle

\begin{abstract}
The core assumption behind using mutants as substitutes for real faults is that they represent them faithfully.
In traditional software testing, this representativeness has been empirically studied by examining the behavioral relationship between mutant detection and real fault detection.
In deep learning (DL), mutants are generated by applying mutation operators to either pre-training or post-training artifacts.
However, the realism of mutants in DL remains largely unverified.

To the best of our knowledge, this study presents the first empirical comparison of pre-training and post-training mutation approaches in DL with respect to realism.
We introduce a statistical framework that quantifies realism using two complementary metrics: coupling strength and behavioral similarity to real faults.
We evaluate the framework on four publicly available bug datasets: \textit{CleanML}, \textit{DeepFD}, \textit{DeepLocalize}, and \textit{defect4ML}.
Mutants are generated using state-of-the-art tools representing both mutation approaches.

Results show that pre-training mutants exhibit significantly stronger coupling and higher behavioral similarity to real faults than post-training mutants, indicating greater realism.
Operator-level analysis shows that this advantage is systematic across individual mutation operators and ranks the operators by realism within each approach.
However, the substantial computational cost of pre-training mutation underscores the need for more effective post-training operators that match or exceed the realism of pre-training mutants.


\end{abstract}

\section{Introduction}
\label{sec:introduction}



Deep learning (DL) systems are increasingly deployed in domains where model reliability is critical, which has driven extensive research on testing and debugging techniques for DL. Many of these techniques rely on mutation analysis to create synthetic faulty versions of models, training scripts, or data for evaluation purposes. These mutants often serve as substitutes for real faults. However, whether these mutants faithfully resemble real DL faults remains unclear, which raises concerns about the validity of mutation-based evaluations. At the same time, real faults are difficult to obtain at scale: \cite{morovati_bugs_2023} report that only 3.48\% of GitHub issues and 2.93\% of Stack Overflow posts involving DL faults were reproducible, and \cite{jahangirova_real_2024} were able to reproduce only 52\% of faults in curated datasets.

The limited availability and reproducibility of real faults in DL research have made mutation testing a common substitute. In the traditional software development paradigm, mutation testing \citep{papadakis_chapter_2019} tools provide an alternate means to generate faulty program versions by injecting small syntactical changes. These modified versions, called mutants, are often used as proxies for real faults. The primary motivation for using mutants as substitutes for real faults lies in their perceived realism. The theory of the realism of mutants in mutation testing is grounded in two foundational hypotheses: the competent programmer hypothesis, and the coupling effect \citep{demillo_hints_1978}. On one hand, the competent programmer hypothesis assumes that programmers usually write nearly correct code. This means that small syntactic changes introduced by mutation operators can reflect developers’ mistakes, as explored in our recent study \citep{ahmed_new_2024}. On the other hand, \cite{demillo_hints_1978} stated the coupling effect as: \textit{``Test data that distinguishes all programs differing from a correct one by only simple errors is so sensitive that it also implicitly distinguishes more complex errors"}. This implies that mutants are useful when they are coupled to real faults, as test inputs that kill the mutant may also detect the corresponding real fault. Empirical evidence from traditional software development strongly supports this assumption, with studies reporting high coupling rates between mutants and real bugs across diverse datasets and environments \citep{just_are_2014, petrovic_does_2021, laurent_re-visiting_2022}. 

Recently, mutation analysis has been adapted for the machine learning paradigm \citep{ma_deepmutation_2018, shen_munn_2018, hu_deepmutation_2019, humbatova_deepcrime_2021, kim_muff_2025}. Beyond test assessment, mutation analysis in DL has been employed across a broad range of tasks including automated program repair \citep{wu_mutation-based_2021, sohn_arachne_2023}, input data prioritization for labeling \citep{wang_prioritizing_2021}, robustness evaluation of models \citep{hu_deepmutation_2019, lin_robustness_2022, mendez_testing_2024}, evaluation of data cleaning techniques \citep{li_cleanml_2021, abdelaal_rein_2023}, generation and detection of adversarial examples \citep{wang_adversarial_2019, pour_search-based_2021}, test data generation \citep{riccio_deepmetis_2021, deokuliar_improving_2023}, oracle generation for autonomous vehicles \citep{jahangirova_quality_2021}, fault localization in deep neural networks \citep{cao_deepfd_2022, ghanbari_mutation-based_2023}, and supporting metamorphic testing strategies in machine learning \citep{xie_testing_2011}. This wide range of use cases underscores the importance of ensuring that DL mutants behave realistically. 

The realism of mutants has far-reaching implications for the broader DL engineering ecosystem. If mutants deviate substantially from the behavior of real faults, these downstream techniques risk drawing misleading conclusions.
Establishing empirical evidence on the realism of DL mutants is therefore essential to ensure that mutation-based evaluation and analysis methods remain credible across diverse applications.
Unlike traditional software systems, mutant detection in DL is often probabilistic due to inherent nondeterminism.
Therefore, the fundamental hypotheses of classical mutation testing do not directly translate to this new data-driven development paradigm \citep{panichella_what_2021}.
We need to carefully reconsider their validity.
In DL, mutants can be generated either via pre-training or post-training approach~\citep{ma_deepmutation_2018, ahmed_exploring_2024}.
The design of pre-training operators~\citep{humbatova_deepcrime_2021} based on real faults offers indirect theoretical support for their realism.
However, design intuition alone does not establish empirical behavioral validity: whether this assumed advantage actually materializes, how large it is, and how consistently it holds across faults and operators remain open questions.
Moreover, pre-training tools such as DeepCrime~\citep{humbatova_deepcrime_2023} generate first-order mutants, each involving a single operator and configuration, whereas real DL faults often include multiple co-occurring issues spanning data preprocessing, architecture, and training configuration. Whether such first-order mutants behaviorally couple to composite real faults cannot be established from operator design alone.
Accordingly, comparing pre-training and post-training mutants against real faults is an empirical question.
To date, no prior work has systematically investigated the extent to which DL mutants represent real faults.

In this study, we aim to quantify and compare the degree of realism\footnote{“Realism” in our context refers to the degree to which a mutant’s test detectability resembles that of its corresponding real fault.} in pre-training and post-training mutants of DL systems.
We propose a statistical method to measure the realism via coupling strength and behavioral similarity between the mutants and the real faults.
To account for nondeterminism, each mutant model is trained multiple times, and evaluated on every test input to compute per-input killing probabilities.
Based on these probabilities, we quantify the realism in terms of the coupling strength and behavioral similarity of each mutant with its corresponding real fault.

This paper makes the following key contributions:
\begin{itemize}
    \item We curate a dataset of reproducible real faults in DL systems, selected from multiple public datasets. We systematically filter out crash-inducing bugs and retain only those faults that (i) exhibit observable behavioral changes, (ii) include test or validation data, and (iii) can be executed in a unified, controlled environment. All bugs were syntactically adapted where necessary to ensure compatibility and reproducibility.
    
    \item We propose a statistical methodology to quantify the relationship between DL mutants (pre-training and post-training) and real faults, enabling a rigorous assessment of mutant representativeness. Our approach introduces two complementary metrics to evaluate the coupling strength and behavioral similarity in a nondeterministic DL context.

    \item We provide the first empirical comparison of pre-training and post-training mutation behavioral realism in DL, analysed at both the approach level and the individual operator level.
    The approach-level analysis quantifies the overall difference between the two paradigms, while the operator-level analysis ranks operators by realism and confirms that the observed trends are systematic rather than driven by a few high-performing operators.

\end{itemize}

The remainder of this paper is organized as follows.
Section \ref{sec:background} provides background on mutation testing for DL.
Section \ref{sec:related-work} summarizes related work.
Section \ref{sec:approach} presents our proposed approach to the problem.
Section \ref{sec:experiments} describes the evaluation methodology used in this study.
Section \ref{sec:results} reports the experimental results, followed by a discussion in Section \ref{sec:discussion}.
Finally, Section \ref{sec:conclusion} concludes the paper and outlines directions for future work.

\section{Background}
\label{sec:background}
This section outlines how mutation testing has been adapted to DL, emphasizing the main mutation targets and approaches.

\subsection{Mutation Testing for Deep Learning}
\label{subsec:mutation4DL}
Mutation analysis has emerged as a crucial methodology for assessing the quality and robustness of DL models \citep{ma_deepmutation_2018, shen_munn_2018, hu_deepmutation_2019, humbatova_deepcrime_2021, kim_muff_2025}. Unlike traditional software development, where logic is manually coded, DL models learn their behavior from data through a training process. This fundamental difference introduces unique artifacts in the DL pipeline where synthetic faults can be injected using mutation operators. Specifically, mutants can be generated by introducing faults into the training program, the training data, or the resulting trained model \citep{ahmed_exploring_2024}. DL model mutation artifacts are broadly categorized into two types: pre-training artifacts such as training program and training data, and the post-training artifact such as trained model.

\subsubsection{Pre-training Approach}
\label{subsubsec:pre-training}
\begin{figure*}
  \includegraphics[width=0.95\textwidth]{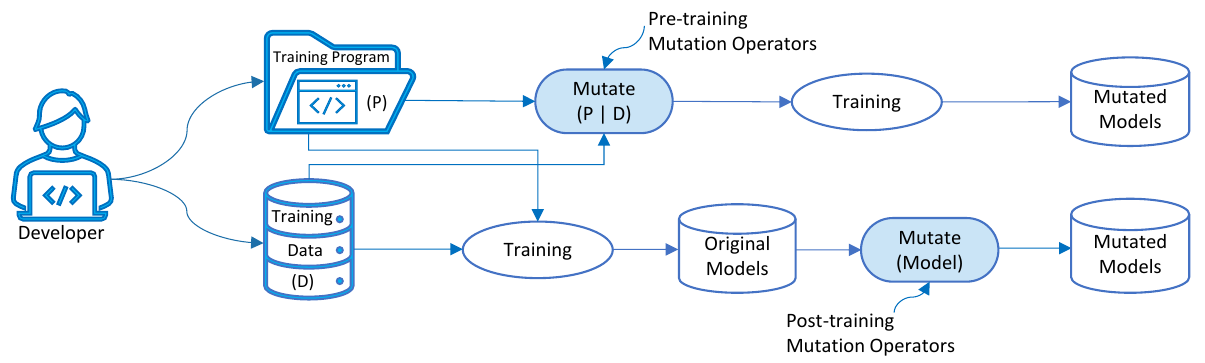}
\caption{Mutant generation with pre-training and post-training approaches}
\label{fig:mutant-generation}       
\end{figure*}

Pre-training mutation analysis involves introducing faults into the data or training scripts prior to the training phase, as illustrated in the upper part of Fig. \ref{fig:mutant-generation}. Such mutations reflect potential faults that could originate during data preparation or training program implementation, thereby generating mutant models through retraining.

DeepMutation \citep{ma_deepmutation_2018} is a pioneering framework that proposes pre-training mutation operators. It defines two categories of pre-training mutation operators: data mutation operators and training program mutation operators. Data mutation operators manipulate training datasets by duplicating data, introducing labeling errors, removing data points, shuffling the data order, or adding noise. Program mutation operators alter the training scripts, such as removing layers, adding layers, or removing activation functions. Mutants are then produced by retraining models on these modified datasets or programs, enabling systematic test data quality evaluation by measuring the ability of test sets to detect behavioral differences between mutants and the original model. Building upon this foundation, DeepCrime \citep{humbatova_deepcrime_2021} substantially enhanced the realism of pre-training mutation operators. \cite{humbatova_deepcrime_2021} systematically derived mutation operators from empirical studies of real faults \citep{zhang_empirical_2018, islam_comprehensive_2019, humbatova_taxonomy_2020} identified through Stack Overflow posts, GitHub issues, and expert interviews. They proposed 35 mutation operators, with 24 implemented at the source level (pre-training). Mutation operators reflected realistic faults such as incorrect labeling, missing data, inappropriate hyperparameter settings (e.g., learning rate, number of epochs), and data imbalance issues. 

Despite their improved realism and sensitivity, pre-training mutants are computationally expensive because each requires a full retraining of the model, greatly increasing computational cost.

\subsubsection{Post-training Approach}
\label{subsubsec:post-training}
In contrast, post-training mutation analysis directly injects faults into already trained models, bypassing the computationally intensive retraining phase associated with pre-training mutant generation, as depicted in the lower part of Fig. \ref{fig:mutant-generation}. DeepMutation \citep{ma_deepmutation_2018} and MuNN \citep{shen_munn_2018} independently pioneered model-level mutation analysis for DL, defining post-training operators such as neuron blocking, activation inversion, weight and bias mutations, and layer manipulations to simulate faults in trained models. These model mutations enabled efficient mutant generation and were the first attempt to introduce systematic mutation analysis into already trained models without additional training. Building upon the concept of model-level mutation operators, DeepMutation++ \citep{hu_deepmutation_2019} extends the original DeepMutation framework \citep{ma_deepmutation_2018}. It retains the eight post-training operators for feedforward neural networks and introduces nine additional operators specifically designed for recurrent architectures, targeting mutations at the weight, state, and gate levels. Addressing the limitations observed in previous post-training approaches, MuFF \citep{kim_muff_2025} introduced two novel mutation operators, Weight Inhibitor and Neuron Inhibitor, alongside an automatic stability-checking mechanism.

Recent empirical studies find that the post-training mutation approach is substantially more computationally efficient than its pre-training counterpart \citep{abbasishahkoo_teasma_2024, kim_muff_2025}. However, efficiency does not imply realism. Faults injected into trained models may not capture the nature of real developer mistakes.

\section{Related Work}
\label{sec:related-work}
In traditional software testing, the coupling between mutants and real faults has been empirically studied by examining the behavioral correlation between mutant detection and real fault detection.
\cite{just_are_2014} showed that mutation score is significantly correlated with real fault detection effectiveness across multiple Java projects, establishing that mutants serve as valid substitutes for real faults in test adequacy assessment.
\cite{petrovic_does_2021} confirmed this relationship at industrial scale, demonstrating that tests generated to kill mutants also detect real bugs in production systems at Google.
\cite{laurent_re-visiting_2022} revisited the coupling between mutants and real faults using the expanded Defects4J~2.0 benchmark and reported that high coupling rates persist across a broader set of projects and fault types.
Across these studies, realism is consistently operationalized as a behavioral relationship: whether tests that distinguish mutants from original programs also distinguish faulty programs from original ones.
Our coupling strength metric adapts this established directional notion to the probabilistic DL context, where nondeterminism necessitates continuous killing probabilities rather than binary kill decisions.

In the DL domain, existing work has advanced mutation testing primarily in terms of efficiency and sensitivity, yet none has directly quantified mutant realism in this coupling sense.
\cite{humbatova_deepcrime_2021} introduced a pre-training mutation testing framework called DeepCrime (see Sec. \ref{subsubsec:pre-training}). While not framed explicitly in terms of realism, DeepCrime examined the sensitivity of mutation testing, where sensitivity refers to the relative variation in the mutation score when the quality of the test set changes. Their study showed that DeepCrime achieves higher sensitivity than DeepMutation++~\citep{hu_deepmutation_2019}. The authors attribute this higher sensitivity to two factors: DeepCrime’s mutation operators are derived from real DL faults, and the mutations are applied before training, allowing the faults to influence the learning process. These observations offer indirect support for the potential realism of pre-training mutants.

In contrast, \cite{kim_muff_2025} proposed an effective post-training mutation framework, MuFF. It emphasizes the efficient generation of stable, sensitive mutants and, in head-to-head comparisons, achieves approximately 61× faster mutant generation than DeepCrime’s pre-training approach \citep{humbatova_deepcrime_2021}. While MuFF does not explicitly assess realism, it evaluates sensitivity to test quality and shows that its mutants are more stable and more computationally efficient than those from DeepCrime, while maintaining comparable (though slightly lower) sensitivity. This trade-off between efficiency and sensitivity positions MuFF as a practical alternative, yet it leaves open whether its mutants reflect the kinds of mistakes real developers make.

Finally, TEASMA~\citep{abbasishahkoo_teasma_2024} explored the fault detection rate of test sets using existing mutation operators from DeepCrime and DeepMutation++. Its relevance to this study lies in its empirical evidence comparing the efficiency of the two approaches: it reports that post-training mutations require about 150× less time than pre-training ones. However, this study does not address the realism, sensitivity, or quality of the mutants used in its adequacy assessment.

In summary, existing studies have advanced DL mutation testing mainly in terms of efficiency and sensitivity, yet none has directly quantified mutant realism across pre-training and post-training paradigms. This gap motivates our study, which empirically compares both approaches to determine how closely their mutants resemble real DL faults and to inform future operator design and evaluation.

\section{Approach}
\label{sec:approach}

The goal of this study is to determine which mutation approach, pre-training or post-training, yields more realistic mutants. To this end, we conduct an empirical evaluation guided by two research questions (RQs) aimed at assessing the realism and representativeness of DL mutants. This section first presents our RQs and their motivations, followed by the detailed methodology we employ to address each question.

\subsection{Research Questions}
\label{sec:research-questions}

Our study is structured around the following research questions (RQs):
\begin{itemize}

    \item \textbf{RQ\textsubscript{1}:} \textit{Do pre- and post-training mutant-killing test inputs also detect real faults?}  
    This question evaluates the coupling strength between mutants and real faults, examining whether tests that detect mutants also expose real faults. By comparing pre- and post-training mutants, we assess whether one mutation approach provides stronger coupling to real faults.

    \item \textbf{RQ\textsubscript{2}:} \textit{How similar is the detectability of pre- and post-training mutants and real faults?}  
    This question investigates the behavioral overlap between mutants and real faults by measuring how similarly they respond to the same test inputs. We compare this overlap across pre-training and post-training mutants to determine which approach more closely resembles real faults in terms of detectability patterns.

\end{itemize}

Both RQs are addressed at two levels of granularity: the approach level, comparing pre-training against post-training across all bugs, and the operator level, examining whether approach-level trends hold uniformly across individual mutation operators.
To systematically address these RQs, we develop a statistical methodology described in the following section.

\subsection{Methodology}
\label{sec:methodology}

In this section, we present our method to quantify the realism of DL mutants. Specifically, we investigate whether pre-training or post-training mutants more closely resemble real faults, which is grounded in the foundational assumption of mutation analysis known as the \textit{Coupling Effect} \citep{demillo_hints_1978}. This effect implies that if mutants are valid substitutes for real faults, then test cases that kill mutants should also detect real faults. Our approach is designed not only to compare the representativeness of pre-training and post-training mutants, but also to quantify how strongly the detection behavior of mutants aligns with that of real faults.

\begin{figure*}
  \includegraphics[width=1.0\textwidth]{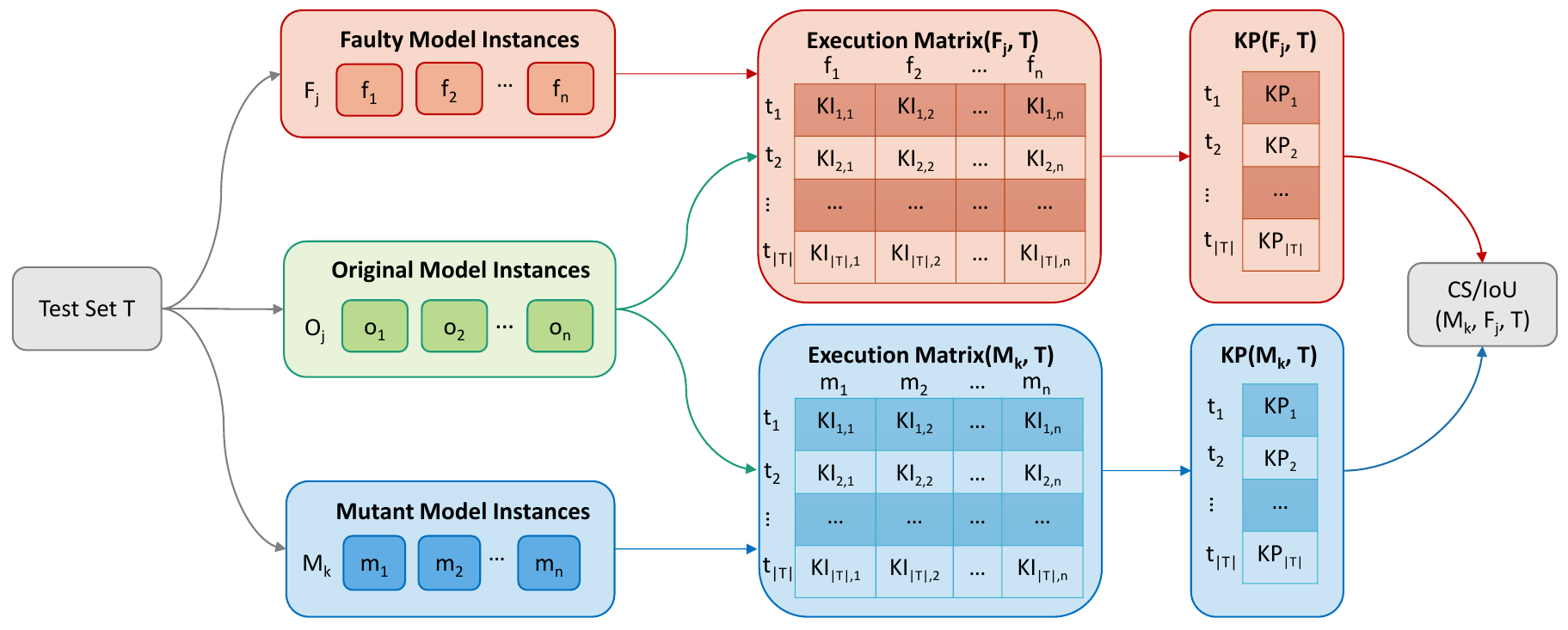}
  \caption{An overview of the proposed methodology. KI, KP, CS, and IoU correspond to the definitions provided in Equations~(\ref{eq:killing-input})--(\ref{eq:iou}).}
  \label{fig:methodology}
\end{figure*}

To this end, we propose a statistical method to compute the realism of DL mutants with respect to real faults, using Coupling Strength (CS) to measure directional coupling between mutants and real faults, and Intersection over Union (IoU) to measure their symmetric behavioral similarity.
Both metrics are derived from the standard Jaccard similarity and containment formulations \citep{broder_resemblance_1997, castro_fernandez_lazo_2019}. 
Starting with a dataset of real bugs and corresponding fixes, we generate mutants from the fixed versions using both pre-training and post-training mutation approaches. 
To account for the stochasticity inherent in DL training, we train each model multiple times, obtaining $n$ independently trained instances per configuration of mutation operators, and construct an \textit{execution matrix} over these instances. 
This matrix captures a binary outcome for each test input, indicating whether it is correctly predicted by the original model instance but incorrectly predicted by a faulty or mutant model.
From this, we derive the \textit{killing probability} \citep{humbatova_deepcrime_2021} of each test input for each mutant. Finally, based on these probabilities, we compute \textit{CS} and \textit{IoU} to analyze the realism of mutants.
Figure~\ref{fig:methodology} summarizes our methodology. We now describe each step of this process in detail.

\subsubsection{Model Training}
\label{sec:model-training}

Our approach involves training three distinct categories of DL models, described as follows:

\begin{itemize}
    \item \textbf{Faulty models} contain real faults collected from existing datasets.
    \item \textbf{Original models}\footnote{“Original model” follows the terminology used by the DeepCrime and DeepMutation++ studies for the unmutated baseline model.} are the fixed versions of the faulty models, representing intended behavior.
    \item \textbf{Mutant models} are synthetically modified versions of the original models.
\end{itemize}

Original and faulty models are trained using the same datasets and hyperparameters, differing solely in the presence or absence of real faults. Mutant models are generated from original models using both pre-training (see Sec.~\ref{subsubsec:pre-training}) and post-training mutation operators (see Sec.~\ref{subsubsec:post-training}).

To account for inherent non-determinism in DL training, each model type is trained multiple times (\textit{n} instances). These repeated instances provide statistical robustness and form the basis for subsequent execution matrix construction.

We denote the set of trained instances as follows: $F = \{f_1, f_2, \dots, f_n\}$ for the faulty models, $O = \{o_1, o_2, \dots, o_n\}$ for the original models, and $M = \{m_1, m_2, \dots, m_n\}$ for the mutant models. These notations are used throughout the subsequent steps of our approach.

\subsubsection{Execution Matrix}
\label{sec:execution-matrix}

To assess the behavioral deviation of both mutant and faulty models from the original model, we construct an \textit{execution matrix} that captures the response of each model instance to every test input. This matrix serves as the foundation for quantifying the extent to which each test input distinguishes mutated or faulty models from correct ones.

Given a test set $T$, we compare each mutant instance $m_i \in M$ and faulty instance $f_i \in F$ with the corresponding original instance $o_i \in O$.

Following the statistical mutant‐killing principle introduced by \citet{jahangirova_empirical_2020} and its probabilistic extension by \citet{humbatova_deepcrime_2021}, a test input $t \in T$ is considered a \textit{killing input (KI)} for a model instance $m_i$ (or $f_i$) if $t$ is correctly classified by $o_i$ but misclassified by $m_i$ (or $f_i$). To record this behavior, we define the binary killing function as:

\begin{equation}
\label{eq:killing-input}
KI_{x_i}(t) = 
\begin{cases}
1, & \text{if } o_i(t) = y \land x_i(t) \neq y \\
0, & \text{otherwise}
\end{cases}
\end{equation}

\noindent where $x_i$ can refer to either $m_i$ or $f_i$, and $y$ is the ground-truth label. A value of 1 indicates that test input $t$ distinguishes model $x_i$ from the original model $o_i$.

The result is a binary execution matrix of size $n \times |T|$ for each comparison (mutants and fault), where each row corresponds to a test input and each column to a model instance. This matrix is then used to compute the killing probability of each test input, described in the next section (see Sec.~\ref{sec:killing-probability}). Although this study focuses on classification tasks, this definition can be extended to regression tasks using threshold-based criteria, as discussed by \citet{humbatova_deepcrime_2021}.

\subsubsection{Killing Probability}
\label{sec:killing-probability}

The execution matrix constructed in the previous step allows us to quantify the likelihood that a test input detects a behavioral difference between a faulty or mutant model and its corresponding original model. This likelihood is formalized as the \textit{killing probability (KP)} of a test input.

Following the probabilistic formulation proposed by \citet{humbatova_deepcrime_2021}, which generalizes the statistical killing notion of \citet{jahangirova_empirical_2020}, the killing probability of a test input $t \in T$ for a set of instances of the model $X = \{x_1, x_2, \dots, x_n\}$ (where $X$ may be either mutant $M$ or faulty model $F$) is computed as:

\begin{equation}
\label{eq:killing-probability}
KP_X(t) = \frac{1}{n} \sum_{i=1}^{n} KI_{x_i}(t)
\end{equation}

\noindent where $KI_{x_i}(t)$ is defined in Equation~\ref{eq:killing-input}.

This metric captures the proportion of model instances in which the test input $t$ is a killing input. A value of $KP_X(t) = 1$ indicates that $t$ consistently exposes behavioral differences across all instances of $X$ when compared to their corresponding original models, while $KP_X(t) = 0$ indicates no such difference was observed in any instance.

We compute separate killing probabilities for mutant and faulty model sets to enable downstream analysis of CS and IoU.
By construction, the downstream comparison metrics give smaller weight to test inputs with lower killing probabilities.
Consequently, infrequent distinctions are retained in the behavioral profile but have a correspondingly smaller effect on the overall comparison.

\subsubsection{Coupling Strength}
\label{sec:coupling-strength}
We first measure how well a mutant mimics a real fault by computing the \textit{CS}. This metric quantifies the proportion of mutant-killing inputs that also detect the faulty model.

For each mutant–fault pair \( (M_k, F_j) \), we define the CS over test set \( T \) as:

\begin{equation}
\label{eq:coupling-strength}
CS(M_k, F_j, T) = \frac{\sum\limits_{t \in T} \min(KP_{M_k}(t), KP_{F_j}(t))}{\sum\limits_{t \in T} KP_{M_k}(t)}
\end{equation}

This formulation corresponds to a weighted Jaccard containment, emphasizing directional influence from mutants to real faults.
A CS of 1 indicates that every test input which detects the mutant also detects the corresponding faulty model with at least the same probability.
This means that the detectability of the mutant is fully contained within that of the faulty model, indicating a high degree of coupling.
This formulation captures directional containment of detectability rather than exact equivalence between the two models.
Lower values suggest weaker coupling and reduced realism.

\subsubsection{Behavioral Similarity}
\label{sec:behavioral-similarity}
In addition, to capture how similarly a mutant and a real fault behave under the same test inputs, we compute the \textit{behavioral similarity} in terms of detectability overlap between their killing probability distributions. This is formalized using an adapted version of the \textit{IoU} metric, also known as the \textit{Jaccard Similarity}, which is commonly used to measure similarity between sets.

Let $KP_{M_k}(t)$ and $KP_{F_j}(t)$ denote the killing probabilities of test input $t$ for a mutant $M_k$ and a faulty model $F_j$, respectively. Given a test set $T$, the IoU between $M_k$ and $F_j$ is computed as:

\begin{equation}
\label{eq:iou}
IoU(M_k, F_j, T) = 
\frac{\sum_{t \in T} \min(KP_{M_k}(t), KP_{F_j}(t))}
     {\sum_{t \in T} \max(KP_{M_k}(t), KP_{F_j}(t))}
\end{equation}

This metric ranges from 0 to 1, with higher values indicating greater overlap in the killing probability distributions of the mutant and the real fault across the test inputs.

\section{Experiments}
\label{sec:experiments}

We conducted an experiment using publicly available DL fault datasets to gain insights into the realism of mutants. In the following, we describe data collection, mutation operators, and experimental setup. Our implementations and results are publicly available.\footnotemark
\footnotetext{https://github.com/zaheedahmed/dl-mutant-realism}

\subsection{Data Collection}
\label{sec:data-collection}

The foundation of our experiments is the dataset of real faults. We curated a benchmark of bug-fix pairs in DL systems from publicly available datasets. Our objective was to collect executable, reproducible, and semantically meaningful real faults affecting either the training code or the data used to train DL models.

\subsubsection{Dataset Identification}
We began by identifying candidate datasets of real faults in DL systems from a recent systematic literature review by ~\citet{jahangirova_real_2024}, which cataloged publicly available DL fault benchmarks and assessed their reproducibility.
We independently verified the reproducibility of each candidate dataset for our experimental setting.
An additional dataset, REIN~\citep{abdelaal_rein_2023}, was identified through a manual search of recent publications on real DL data faults. This resulted in an initial pool of seven publicly available and potentially reproducible datasets (see Sec.~\ref{sec:dataset-analysis}).
Each dataset was analyzed based on its accompanying publication and available replication package.

\subsubsection{Dataset Analysis}
\label{sec:dataset-analysis}

The seven datasets fall into two distinct categories, \textit{data faults} and \textit{program faults}, based on the origin of the faults. This distinction guided how we examined and selected faults for inclusion:

\begin{itemize}
\item \textbf{Data Faults:} These faults stem from issues in the training data, such as missing values, mislabeled entries, outliers, or duplicates. CleanML~\citep{li_cleanml_2021} provides faulty datasets and several cleaning algorithms but no definitive fixed version. Since it uses classical machine learning classifiers, we added a unified Keras-based multi-layer perceptron (MLP) to ensure compatibility with DL mutation testing frameworks. For each fault, we applied all relevant cleaning methods to the faulty dataset, trained the MLP with default hyperparameters on each faulty and fixed version, and selected the fixed dataset with the highest performance. This process yielded a bug–fix pair for every fault, consisting of the faulty dataset, its best fixed counterpart, and the MLP. In the CleanML GitHub repository, two datasets listed in the original paper were missing, one unlisted dataset was included, and four synthetic faults were excluded, resulting in 19 real faults across 13 datasets. In REIN~\citep{abdelaal_rein_2023}, only two of the 14 datasets contain real faults; one was unavailable due to a broken URL, leaving only the Beer~\citep{hould_craftcans_2017} dataset accessible. Because it represents just a single instance, we excluded it from our study.

\item \textbf{Program Faults:} These faults originate from the training code (e.g., incorrect layer configurations, misuse of activation or loss functions, or faulty data-preprocessing logic). Each dataset includes faulty and fixed versions of the training scripts with their associated data. We selected faults based on script executability, classification task type, availability of test/validation data, use of Keras/TensorFlow via the Sequential or Functional API, and compatibility with Python 3.8. To ensure a single, reproducible environment, we standardized all scripts originally targeting Python 3.6-3.8 to Python 3.8 by applying only non-semantic, structural updates to deprecated APIs (e.g., \texttt{nb\_epoch}→\texttt{epochs}, \texttt{lr}→\texttt{learning\_rate}) and verified that these changes did not affect model performance beyond expected stochastic variation. Faults were excluded if they crashed irrecoverably, lacked evaluation data, or involved regression or other unsupported tasks. The SFData dataset~\citep{wu_automatically_2022} was excluded due to its exclusive focus on tensor-shape faults, which produce non-executable faulty models, while TFBugs~\citep{zhang_empirical_2018} was excluded because it was the oldest among all and its faults were in old version of Python that require more effort to adapt them for unified environment. This standardization yields a curated, unified corpus of bug-fix pairs executable under a single environment.
\end{itemize}

Table~\ref{tab:datasets-analysis} summarizes the datasets included in our analysis by listing their fault origins and showing how many faults from each dataset contribute to our final benchmark.
Table~\ref{tab:fault-categories} presents the fault category distribution of all 86 faults included  in our benchmark across four datasets mapped to the taxonomy of~\citet{humbatova_taxonomy_2020}.
The benchmark covers 4 of 5 top-level categories within the taxonomy, spanning Model, Training, Tensors \& Inputs, and API categories.
Total count exceeds 86 because complex faults are mapped to multiple categories.

The selected faults span a range of training dataset scales, model depths, and fault types. Training set sizes range from 13 to over 1.9 million samples, covering standard benchmarks (e.g., MNIST, CIFAR-10, Fashion-MNIST, IMDB, GTSRB) as well as domain-specific and synthetic datasets. Model depths range from 3 to 29 layers, implemented using the Keras Sequential or Functional API. The benchmark includes both single-fault and composite bugs: faults in DeepFD and DeepLocalize frequently involve multiple co-occurring issues, while those in CleanML and defect4ML are predominantly single-fault. Detailed per-bug descriptive statistics are provided in Appendix Tables ~\ref{tab:defect4ml-fault-categories}--\ref{tab:cleanml-details}.
As the four datasets span two fault origins and vary in fault complexity, we report results per dataset in Section~\ref{sec:results}.
This allows readers to assess how consistently the observed trends hold across independently curated fault corpora and keeps the per-bug visualisations readable.

\begin{table}
\caption{Summary of selected datasets and faults}
\label{tab:datasets-analysis}
\begin{tabular}{llcc}
\toprule
\textbf{Dataset} & \textbf{Fault Origin} & \textbf{Total Faults} & \textbf{Included Faults} \\
\midrule
CleanML~\citep{li_cleanml_2021}         & Training Data    & 23  & 19 \\
DeepFD~\citep{cao_deepfd_2022}          & Training Program & 58  & 25 \\
DeepLocalize~\citep{wardat_deeplocalize_2021} & Training Program & 40  & 18 \\
defect4ML~\citep{morovati_bugs_2023}   & Training Program & 100 & 24 \\
\noalign{\smallskip}\hline\noalign{\smallskip}
\textbf{Total} & \textbf{---} & \textbf{221} & \textbf{86} \\
\bottomrule
\end{tabular}
\end{table}

\begin{table}[t]
\caption{Fault category distribution across all selected faults mapped to the~\cite{humbatova_taxonomy_2020} taxonomy.}
\label{tab:fault-categories}
\centering
\small
\begin{tabular}{lc}
\toprule
\textbf{Taxonomy Category} & \textbf{\# Faults} \\
\midrule
Model $>$ Model Type \& Properties & 5 \\
Model $>$ Layers $>$ Missing/Redundant/Wrong Layer & 8 \\
Model $>$ Layers $>$ Layer Properties & 9 \\
Model $>$ Layers $>$ Activation Function & 33 \\
\midrule
Training $>$ Hyperparameters & 34 \\
Training $>$ Loss Function & 20 \\
Training $>$ Preprocessing of Training Data $>$ Missing preprocessing & 12 \\
Training $>$ Preprocessing of Training Data $>$ Wrong preprocessing & 2 \\
Training $>$ Optimiser & 6 \\
Training $>$ Training Data Quality & 20 \\
\midrule
Tensors \& Inputs $>$ Wrong Tensor Shape & 1 \\
\midrule
API & 1 \\
\bottomrule
\multicolumn{2}{l}{\footnotesize Note: Complex faults are counted separately for each taxonomy category they trigger.} \\
\end{tabular}
\end{table}

\subsection{Mutation Operators}
\label{sec:mutation-frameworks}

The choice of mutation operators is central to our investigation of mutant realism.
To ensure methodological soundness and practical relevance, we adopt a tool-centric strategy, selecting one state-of-the-art framework for each mutation paradigm: pre-training and post-training.
Our selection criteria emphasize public availability of tool implementations, peer-reviewed publication status, operator diversity, and ease of integration into empirical evaluation pipelines.

For the pre-training approach, we selected DeepCrime~\citep{humbatova_deepcrime_2023}, the only publicly available pre-training mutation tool.
Its mutation operators are explicitly derived from empirical studies of real DL faults~\citep{humbatova_deepcrime_2021}.
For the post-training approach, we selected DeepMutation++~\citep{hu_deepmutation_2019}, a state-of-the-art framework with a publicly available implementation providing 17 mutation operators: eight for feedforward networks and nine for recurrent networks.
Other frameworks, such as DeepMutation~\citep{ma_deepmutation_2018} and MuNN~\citep{shen_munn_2018}, lack public implementations.
MuFF~\citep{kim_muff_2025}, while available, was not included in our evaluation because it offers only two operators and remained an unreviewed preprint at the time of study design.
Table~\ref{tab:mutation-tools} summarises the reviewed frameworks, including their operator scope, implementation status, and selection for our evaluation.

\begin{table}
\centering
\caption{Comparison of DL mutation testing frameworks}
\label{tab:mutation-tools}
\begin{tabular}{l l c c c}
\toprule
\textbf{Name} & \textbf{Operator Type} & \textbf{\# Operators} & \textbf{Tool Available} & \textbf{Selected} \\
\midrule
DeepCrime       & Pre-training       & 24                & \cmark          & \cmark \\
DeepMutation    & Both               & 16                 & \xmark          & \xmark \\
DeepMutation++  & Post-training      & 17                & \cmark          & \cmark \\
MuNN            & Post-training      & 5                 & \xmark          & \xmark \\
MuFF            & Post-training      & 2                 & \cmark          & \xmark \\
\bottomrule
\end{tabular}
\end{table}

\subsubsection{Pre-training Mutation Operators}
\label{sec:pretraining-operators}

Table~\ref{tab:deepcrime_operators} lists the 24 DeepCrime mutation operators used in this study. These operators are organised into eight groups based on their mutation target~\citep{humbatova_deepcrime_2021}: training data (TCL, TRD, TUD, TAN, TCO), hyperparameters (HBS, HLR, HNE, HDB), activation functions (ACH, ARM, AAL), regularisation (RAW, RCW, RRW, RCD, RCP), weights (WCI, WAB, WRB), loss function (LCH), optimisation (OCH, OCG), and validation (VRM). Its fault-informed design and broad operator coverage make it well-suited for evaluating mutant realism in pre-training settings. Operator applicability depends on whether the training script contains the target artifact; for instance, \texttt{Change Dropout Rate} requires a dropout layer in the original model.

\begin{table}
\centering
\scriptsize
\caption{Pre-training mutation operators of DeepCrime.}
\label{tab:deepcrime_operators}
\begin{tabular}{p{0.05\linewidth} p{0.32\linewidth} p{0.65\linewidth}}
\toprule
\textbf{ID} & \textbf{Operator Name} & \textbf{Description} \\
\midrule
TCL & Change labels of training data & Replaces labels of a selected portion of the training data. \\
TRD & Remove portion of training data & Deletes a selected portion of the training data. \\
TUD & Unbalance training data & Removes data from underrepresented classes. \\
TAN & Add noise to training data & Injects noise into a selected portion of the training inputs. \\
TCO & Make output classes overlap & Duplicates one dominant class and relabels it as another. \\
HBS & Change batch size & Replaces the original batch size with a different value. \\
HLR & Decrease learning rate & Lowers the original learning rate. \\
HNE & Change number of epochs & Replaces the original number of training epochs. \\
HDB & Disable data batching & Trains without mini-batches, using the full dataset per update. \\
ACH & Change activation function & Replaces a non-linear activation with a different one. \\
ARM & Remove activation function & Replaces a layer's activation with a linear activation. \\
AAL & Add activation function to layer & Adds a non-linear activation to a layer with linear activation. \\
RAW & Add weights regularisation & Adds kernel regularisation to a layer that originally has none. \\
RCW & Change weights regularisation & Replaces a layer's existing kernel regulariser with a different one. \\
RRW & Remove weights regularisation & Removes kernel regularisation from a layer. \\
RCD & Change dropout rate & Replaces the dropout rate of a dropout layer. \\
RCP & Change patience parameter & Replaces the patience setting of early stopping. \\
WCI & Change weights initialisation & Replaces the kernel initialiser of a layer. \\
WAB & Add bias to a layer & Adds a bias term to a layer without bias. \\
WRB & Remove bias from a layer & Removes the bias term from a layer. \\
LCH & Change loss function & Replaces the training loss function. \\
OCH & Change optimisation function & Replaces the optimiser used for training. \\
OCG & Change gradient clipping & Replaces the gradient clipping value used by the optimiser. \\
VRM & Remove validation set & Removes validation data from the training process. \\
\bottomrule
\end{tabular}
\end{table}

\subsubsection{Post-training Mutation Operators}
\label{sec:posttraining-operators}

Table~\ref{tab:deepmutationpp_operators} lists the eight feedforward neural network operators of DeepMutation++ used in this study.
These operators target three levels of the trained model: weights (GF, WS), neurons (NEB, NAI, NS), and layers (LR, LA, LD). The weight-level and neuron-level operators are parameterized by a \emph{ratio} that controls the proportion of affected weights or neurons.
The layer-level operators perform structural changes (removing, adding, or duplicating a layer) without a ratio parameter.
The remaining nine operators of DeepMutation++ are omitted, as they target recurrent neural networks and are not applicable to any subject system considered in this study.

\begin{table}
\centering
\scriptsize
\caption{Post-training mutation operators of DeepMutation++ considered in this study.}
\label{tab:deepmutationpp_operators}
\begin{tabular}{p{0.05\linewidth} p{0.27\linewidth} p{0.65\linewidth}}
\toprule
\textbf{ID} & \textbf{Operator Name} & \textbf{Description} \\
\midrule
GF  & Gaussian Fuzzing       & Adds Gaussian-distributed noise to a selected portion of weights. \\
WS  & Weight Shuffle         & Shuffles a selected portion of weights within layers. \\
NEB & Neuron Effect Block    & Sets a neuron's output to zero, blocking its effect. \\
NAI & Neuron Activation Inverse & Inverts the sign of a neuron's activation value. \\
NS  & Neuron Switch          & Swaps two neurons within the same layer. \\
LR  & Layer Remove           & Removes one layer from the trained model. \\
LA  & Layer Addition         & Inserts an additional activation layer into the trained model. \\
LD  & Layer Duplication      & Duplicates an existing layer in the trained model. \\
\bottomrule
\end{tabular}
\end{table}

\subsection{Experimental Setup}
\label{sec:experimental-setup}

We designed a unified experimental pipeline to evaluate mutant realism across two fault types: data faults and program faults. Both follow a common structure involving the training of original, mutant, and faulty models, followed by realism analysis.
For each dataset, we first train the original models using the corrected version of both the training program and dataset.
Pre-training mutants are generated with DeepCrime~\citep{humbatova_deepcrime_2023} by injecting faults into the training code or data.
As the majority of DeepCrime operators are parametric, we used its default automated search strategy for parameter selection. DeepCrime applies binary search for range-based parameters (e.g., the percentage of training data to delete) and exhaustive search over list-based parameters (e.g., available activation functions), targeting killable but non-trivial configurations~\citep{humbatova_deepcrime_2021}. No manual parameter overrides were applied.
Post-training mutants are generated by applying DeepMutation++~\citep{hu_deepmutation_2019} to the trained original models.
Most DeepMutation++ operators are parametric, with a ratio parameter that controls the proportion of neurons or weights to mutate.
DeepCrime accounts for training randomness by treating a single mutant as multiple trained instances originating from the same mutated source, whereas DeepMutation++ assumes each mutant is represented by a single instance.
To make the two tools comparable and give DeepMutation++ a fair opportunity to generate diverse mutants, we use two post-training mutation scenarios.
In the first, we apply the default operator configuration multiple times ($n = 5$) to each instance of the original model. In the second, we apply ratio settings from the default to the most aggressive, that is 0.01, 0.1, 0.2, 0.6 and 1.0, once per instance of the original model, following the parameterization strategy of \citet{kim_muff_2025}.
As a result, CS and IoU were computed over all generated mutants in both approaches. For the pre-training approach, non-killable and trivial mutants are excluded through DeepCrime's built-in killability and triviality criteria.
For post-training mutants, DeepMutation++ applies a quality threshold during generation and retains only mutants above the given threshold.
We used the default threshold value of 0.9 for both scenarios.

Across the 86 bugs, DeepCrime generated a median of 153 mutants per bug (min: 2, max: 367), reflecting its automated search across multiple parameter configurations per operator.
The number of mutants per operator varies with the size of its parameter space: operators with list-based parameters (e.g., ACH, WCI) generate tens of configurations, while parameterless operators (e.g., VRM, HDB) generate exactly one.
DeepMutation++ generated a median of 30 mutants per bug under Scenario~1 (min: 18, max: 40) and 16 under Scenario~2 (min: 6, max: 25).
Detailed per-operator counts are available in the replication package.

The generation of faulty models depends on the fault type: (1) for data faults, models are trained using the dirty (faulty) version of the dataset alongside the same training program used for original models; (2) for program faults, models are trained using the faulty version of the training program and fixed dataset.

Each model is trained and evaluated five times ($n = 5$) to account for stochastic variability in DL training.
Prior work on the reproducibility of machine learning experiments has shown that results are often averaged over a small number of multiple independent training runs, typically in the range of 3--10~\citep{agarwal_deep_2021}.
Several studies employ a small number of runs within this range, including configurations with five runs (e.g.,~\cite{henderson_deep_2018}).
Following this established practice, we adopt $n = 5$ as a practical compromise between statistical robustness and computational feasibility.

\begin{figure}
    \centering
    \includegraphics[width=0.95\linewidth]{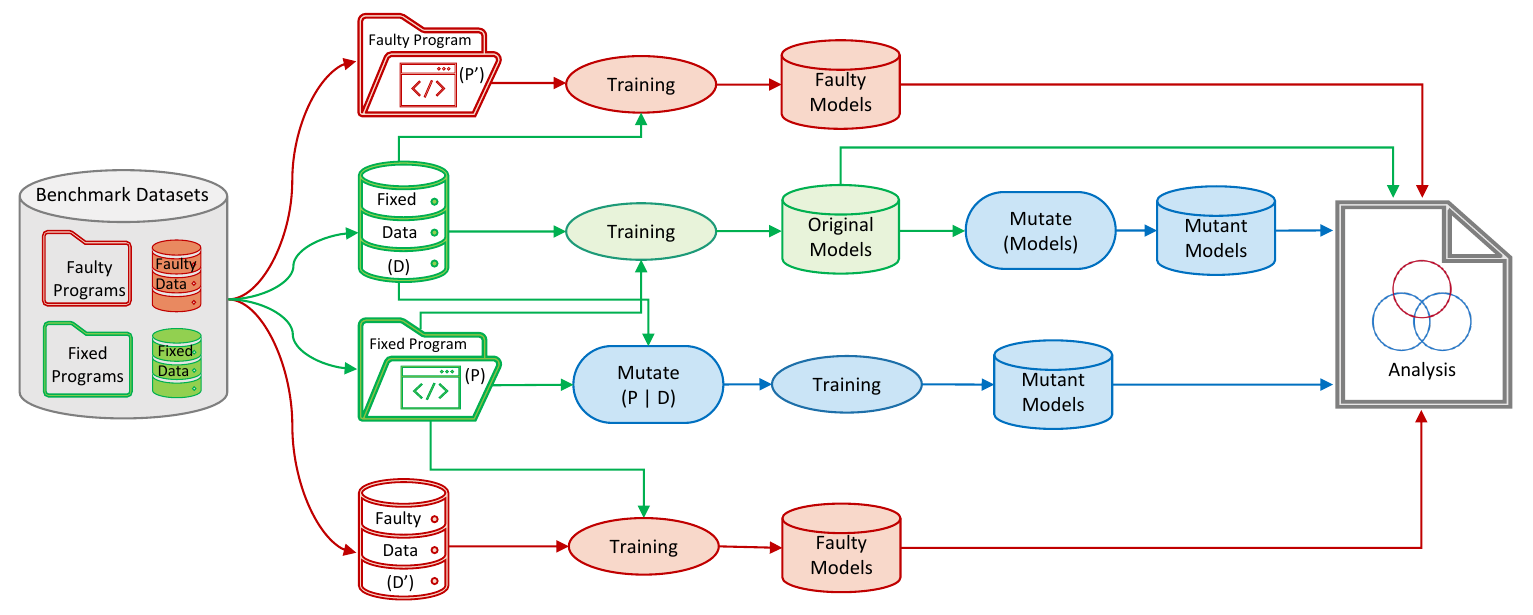}
    \caption{Experimental pipeline for evaluating mutant realism}
    \label{fig:experimental-pipeline}
\end{figure}

This trade-off is particularly important in our setting due to the high cost of mutation analysis.
Pre-training mutant generation required days of computation for several individual subject systems.
Post-training mutation under the aggressive parameterization of Scenario~2 further increased the computational cost, as multiple candidate mutants had to be generated to satisfy quality constraints (e.g., maintaining accuracy above a threshold).
In addition, each mutant model instance was evaluated against the full test set to compute the execution matrix, multiplying the total cost across 86 bugs and three mutation types (pre-training and two post-training scenarios) per bug.
Increasing the number of runs beyond five would lead to a substantial increase in computational cost.
Moreover, the consistency of our results across datasets and scenarios suggests that five runs provide adequate statistical stability for the comparisons reported here.
Figure~\ref{fig:experimental-pipeline} illustrates the full setup.

\paragraph{Tooling and Environment:}
We upgraded DeepMutation++ (originally designed for Python~3.6, TensorFlow~1.13, and Keras~2.2.4) to run on Python~3.8.20, TensorFlow~2.10.0, and Keras~2.10.0, enabling a unified experimental environment. Only API adapters were introduced, and operator semantics remained unchanged.

\subsection{Metrics}
\label{sec:metrics}
To answer our research questions, we use two complementary metrics introduced in Section~\ref{sec:methodology}: \textit{CS} and \textit{IoU}. 

For \textbf{RQ\textsubscript{1}}, we evaluate the degree to which mutant-killing test inputs also detect real faults using the CS metric. A higher CS indicates stronger directional coupling between a mutant and the corresponding faulty model whose fixed version was used to generate that mutant.

For \textbf{RQ\textsubscript{2}}, we assess behavioral similarity between mutants and real faults using IoU, which measures the symmetric overlap in their detectability patterns.

Both metrics range from~0 (no relation) to~1 (perfect correspondence). These two measures jointly capture complementary aspects of mutant realism: CS reflects directional fault substitutability, while IoU reflects behavioral similarity in test response.

\subsection{Statistical Analysis}
\label{sec:statistical_analysis}

To determine whether the observed differences in CS and IoU between pre-training and post-training mutation approaches are statistically significant, we apply the Wilcoxon signed-rank test~\citep{wilcoxon_individual_1945}.
For each bug and metric, we compute the median score across all mutants within each scenario, yielding paired observations across all 86 bugs.
The Wilcoxon signed-rank test is a non-parametric test for paired samples that does not assume normality, making it appropriate for our bounded metrics that range from 0 to 1 and exhibit skewed distributions.

We conduct six planned comparisons derived directly from our research questions.
For each metric, CS and IoU, we compare pre-training against post-training scenario~1, pre-training against post-training scenario~2, and post-training scenario~1 against post-training scenario~2.
All tests are two-sided at a significance level of $\alpha = 0.05$.

To quantify the magnitude of observed differences, we report the rank-biserial correlation $r$~\citep{kerby_simple_2014} as the effect size measure.
The rank-biserial correlation is the standard effect size for the Wilcoxon signed-rank test and is computed directly from the signed ranks.
Following Cohen's conventions~\citep{cohen_statistical_1988}, we interpret the effect size as: negligible ($|r| < 0.1$), small ($0.1 \leq |r| < 0.3$), medium ($0.3 \leq |r| < 0.5$), and large ($|r| \geq 0.5$).

\section{Results}
\label{sec:results}

We present our results organised by the two research questions.
For each question, we first analyse results at the approach level, comparing pre-training against both post-training scenarios across all bugs.
We then drill down to the operator level to examine whether the approach-level trends hold uniformly across individual mutation operators.

\subsection{RQ\textsubscript{1}: Do pre- and post-training mutant-killing test inputs also detect real faults?}

\paragraph{Approach-level Analysis:}
To address RQ\textsubscript{1}, we analyze the \emph{CS} between mutants and real faults across all datasets. CS quantifies the probability that a test input killing a mutant also detects a real fault. 

\begin{figure}[htbp]
  \centering
  \subfloat[CleanML dataset\label{fig:cs_cleanml}]{%
    \includegraphics[width=1.2\linewidth,height=0.25\textheight,keepaspectratio]{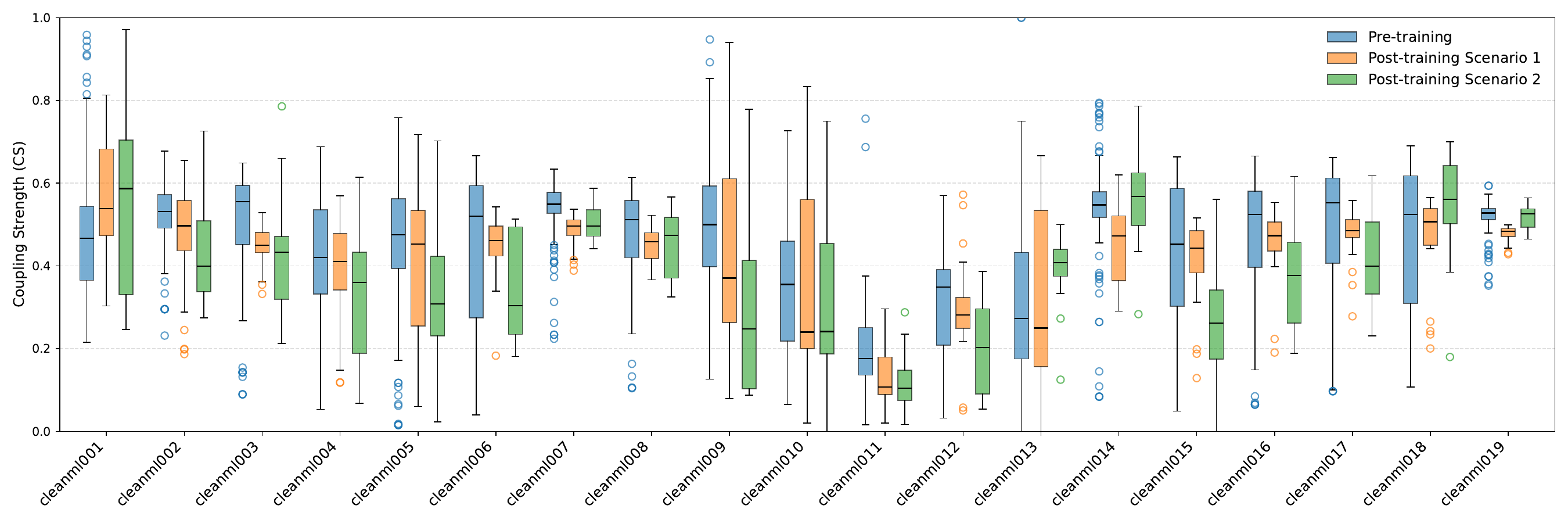}%
  }\\
  
  \subfloat[DeepFD dataset\label{fig:cs_deepfd}]{%
    \includegraphics[width=1.2\linewidth,height=0.25\textheight,keepaspectratio]{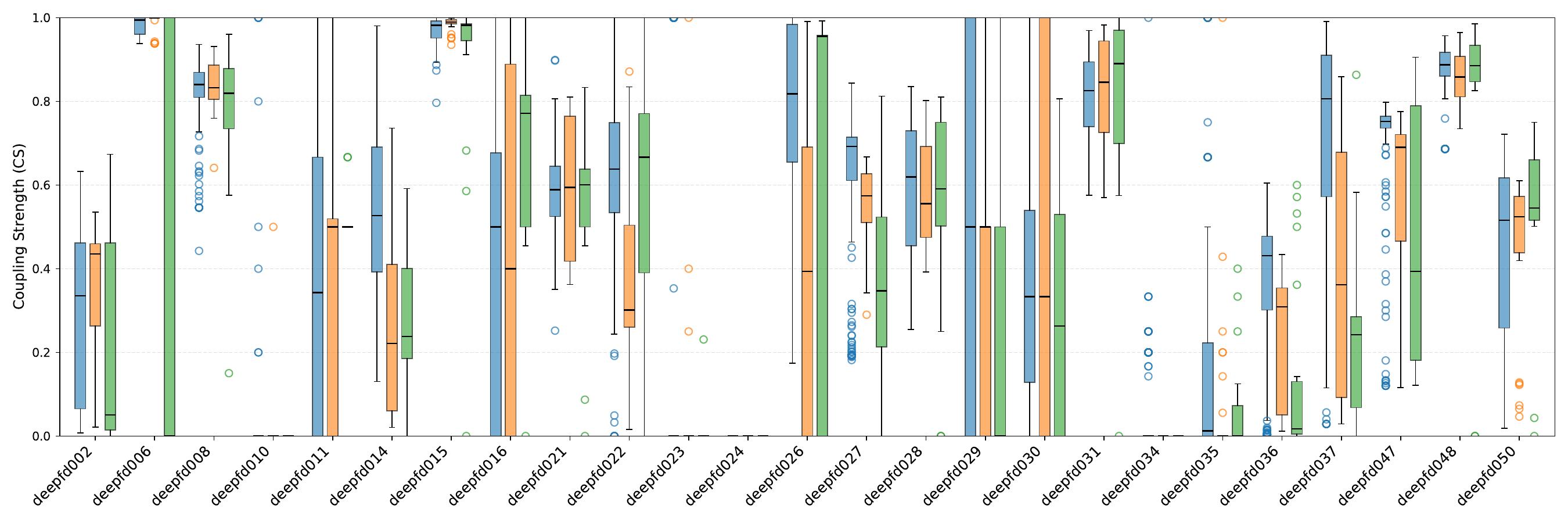}%
  }\\
  \subfloat[DeepLocalize dataset\label{fig:cs_deeplocalize}]{%
    \includegraphics[width=1.2\linewidth,height=0.25\textheight,keepaspectratio]{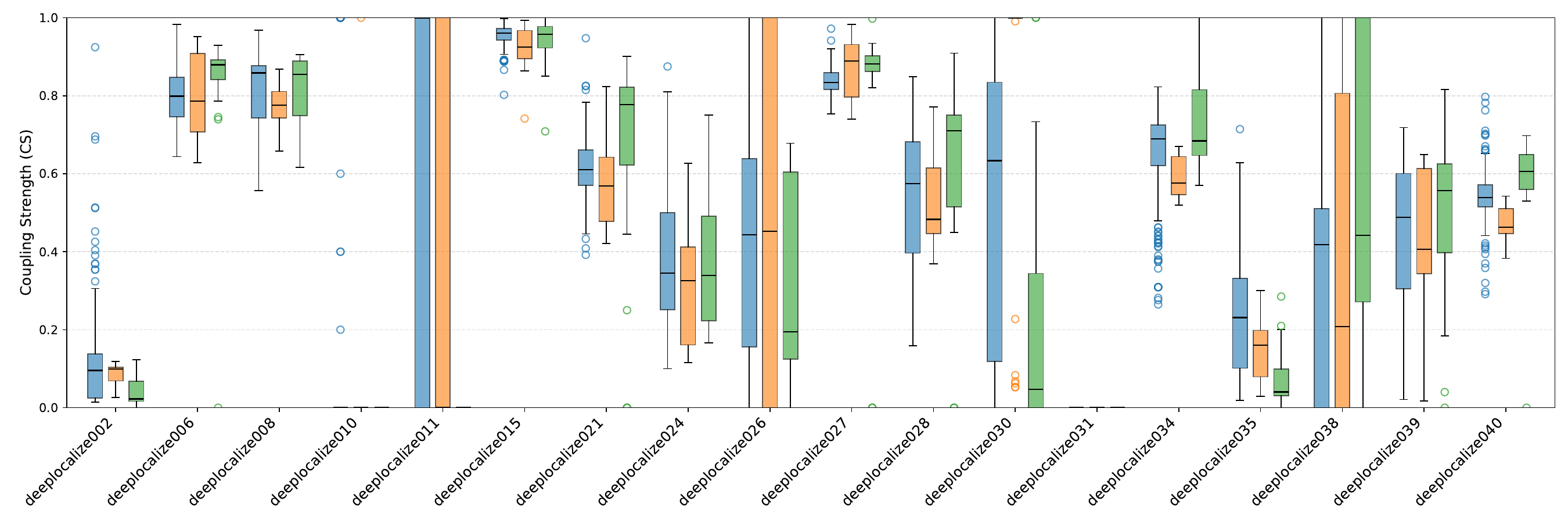}%
  }\\
  \subfloat[defect4ML dataset\label{fig:cs_defect4ml}]{%
    \includegraphics[width=1.2\linewidth,height=0.25\textheight,keepaspectratio]{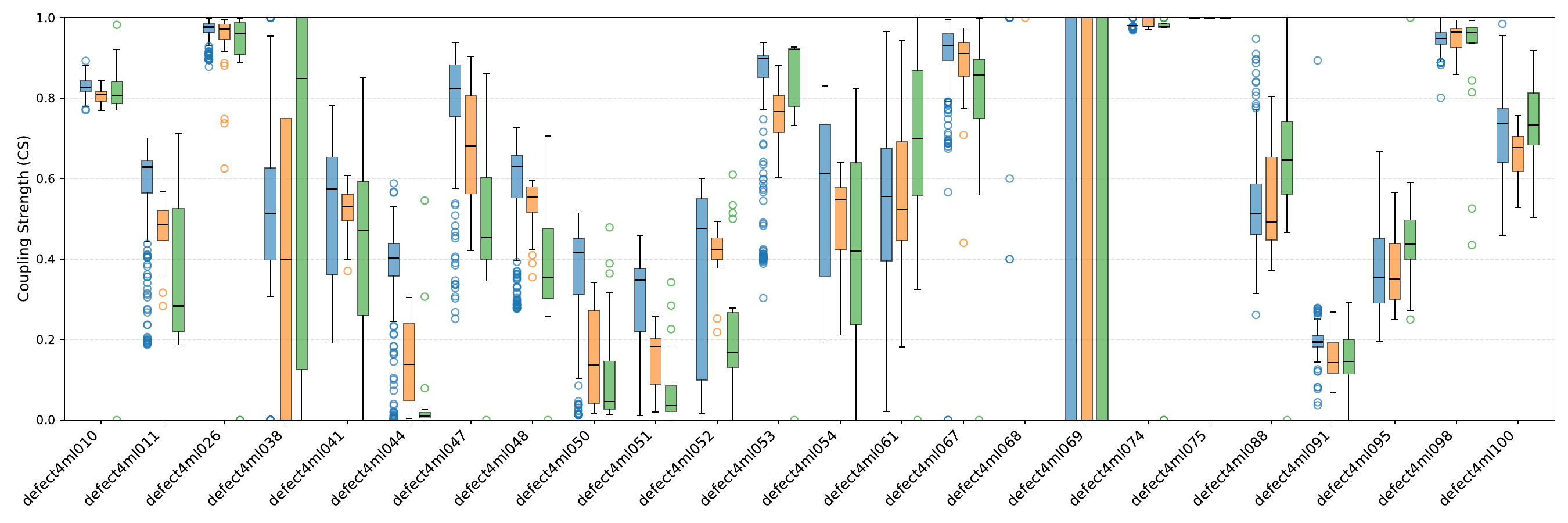}%
  }
  
  \caption{Bug-wise coupling strength distributions across datasets. Each bug shows three scenarios: pre-training, post-training scenario 1, and post-training scenario 2. The color coding and legend shown in subfigure~(a) apply consistently to all subfigures~(b)--(d).}
  \label{fig:cs_boxplots}
\end{figure}

Figure~\ref{fig:cs_boxplots} presents bug-wise distributions of CS values for each dataset: CleanML (Fig.~\ref{fig:cs_cleanml}), DeepFD (Fig.~\ref{fig:cs_deepfd}), DeepLocalize (Fig.~\ref{fig:cs_deeplocalize}), and defect4ML (Fig.~\ref{fig:cs_defect4ml}). Each bug is represented by three boxplots corresponding to \emph{pre-training}, \emph{post-training scenario~1}, and \emph{post-training scenario~2} (see Sec.~\ref{sec:experimental-setup}) mutants. Boxplots reveal not only the medians but also variability and the presence of outliers across mutants. 

\begin{figure}
    \centering
    \includegraphics[width=1.0\linewidth]{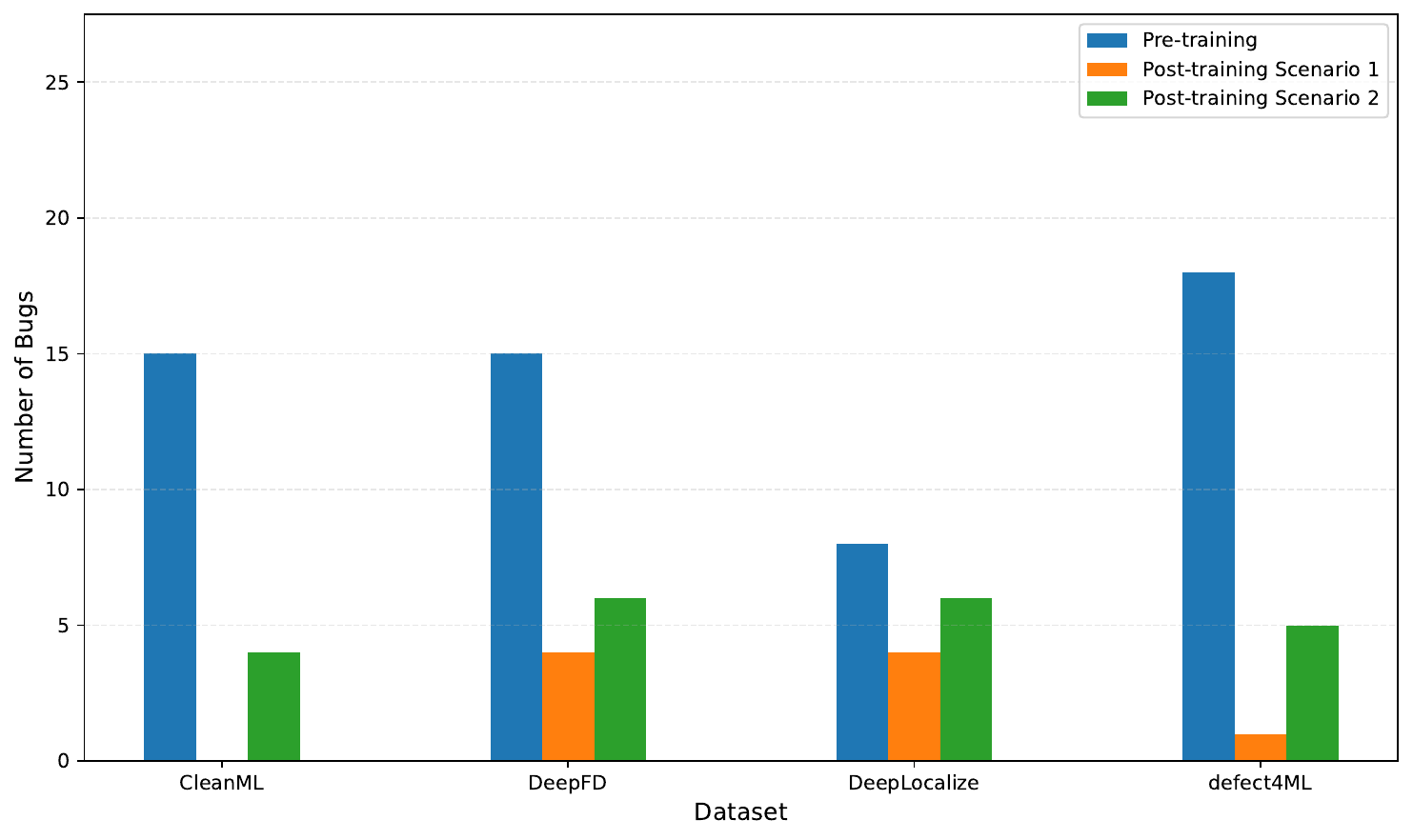}
    \caption{Dataset-level summary of bug-wise cases achieving the highest median coupling strength.}
    \label{fig:win_counts_CS}
\end{figure}

Across datasets, pre-training mutants generally achieve higher median CS values compared to post-training scenarios. 
Notably, the distributions for pre-training mutants also show a higher frequency of upper-tail outliers, indicating the presence of particularly realistic mutants. 
While the median provides a robust measure of central tendency, it does not reflect the presence of these high-performing outliers.
Therefore, when summarizing results at the dataset level based on which scenario attains the highest median CS for each bug (Fig.~\ref{fig:win_counts_CS}), we resolved ties in favor of the pre-training mutation approach.

This tie-breaking strategy acknowledges the greater potential of pre-training mutations to generate highly realistic mutants, as empirically evidenced in Fig.~\ref{fig:cs_boxplots}, where pre-training mutations more frequently yield upper-tail outliers, that is, mutants with exceptionally high coupling to real faults.

Breaking down by dataset, the results in Fig.~\ref{fig:win_counts_CS} show a clear dominance of pre-training mutants: 
\begin{itemize}
    \item \textbf{CleanML (19 data bugs):} Pre-training attains the highest median CS for 15 bugs (79\%), post-training scenario 1 for 0 bugs, and post-training scenario 2 for 4 bugs (21\%).
    \item \textbf{DeepFD (25 program bugs):} Pre-training yields the highest median CS in 15 bugs (60\%), post-training scenario 1 in 4 bugs (16\%), and post-training scenario 2 in 6 bugs (24\%).
    \item \textbf{DeepLocalize (18 program bugs):} Pre-training achieves the highest median CS for 8 bugs (44\%), post-training scenario 1 for 4 bugs (22\%), and post-training scenario 2 for 6 bugs (33\%).
    \item \textbf{defect4ML (24 program bugs):} Pre-training results in the highest median CS for 18 bugs (75\%), post-training scenario 1 for 1 bug (4\%), and post-training scenario 2 in 5 bugs (21\%). 
\end{itemize}

A dataset-level examination confirms that the pre-training advantage holds across all individual datasets.
A notable difference is observed between data and program fault subjects as shown in Fig.~\ref{fig:cs_boxplots}.
CleanML, the data-fault dataset, shows lower absolute CS values for both approaches than the program-fault datasets.
The margin between pre- and post-training is also smaller for data-fault subjects, with fewer upper-tail outliers.
In the program-fault datasets, pre-training mutants more frequently produce outliers approaching CS near 1.0, whereas such near-perfect coupling is substantially less frequent in data-fault subjects.

Aggregating across all datasets (86 bugs in total), pre-training mutants outperform both post-training scenarios in 56 bugs (65\%), while post-training scenario~1 attains the highest median CS in 9 bugs (10\%), and post-training scenario~2 in 21 bugs (24\%).

Table~\ref{tab:cs_statistical} presents the results of the Wilcoxon signed-rank tests on per-bug median CS values across all 86 bugs.
The differences between pre-training and both post-training scenarios are statistically significant ($p < 0.001$), with a large effect size for post-training scenario~1 ($r = 0.758$) and a medium effect size for post-training scenario~2 ($r = 0.467$).
No statistically significant difference was found between the two post-training scenarios ($p = 0.295$, $r = 0.140$, small).

These findings suggest that pre-training mutants serve as more reliable test targets, providing stronger alignment with real faults compared to post-training mutants.
The effect is particularly strong in the program-bug datasets, DeepFD and defect4ML, whereas CleanML, which contains data bugs, shows smaller but still consistent advantages for pre-training. DeepLocalize, another program-bug dataset, presents the most balanced outcome, although pre-training still leads overall.  

Despite the general trend in favor of pre-training, a few exceptions exist. These deviations, while limited in number, indicate that the superiority of pre-training mutants is not universal across all bugs. We revisit these exceptions and discuss potential reasons in Section~\ref{sec:reasons-for-exceptions}.

\begin{table}[t]
\centering
\caption{Statistical comparison of mutation approaches based on coupling strength across all 86 bugs.
Per-bug median CS values are compared using the Wilcoxon signed-rank test (two-sided).
Effect size is measured by the rank-biserial correlation ($r$).}
\label{tab:cs_statistical}
\begin{tabular}{lccccl}
\toprule
\textbf{Comparison} & \textbf{$n$} & \textbf{$p$-value} & \textbf{$r$} & \textbf{Effect} \\
\midrule
Pre-training vs Post-training Scenario 1 & 86 & $<$0.001 & 0.758 & large \\
Pre-training vs Post-training Scenario 2 & 86 & $<$0.001 & 0.467 & medium \\
Post-training Scenario 1 vs Post-training Scenario 2 & 86 &    0.295 & 0.140 & small \\
\bottomrule
\end{tabular}
\end{table}

\begin{figure*}[t]
\centering
\includegraphics[width=\textwidth]{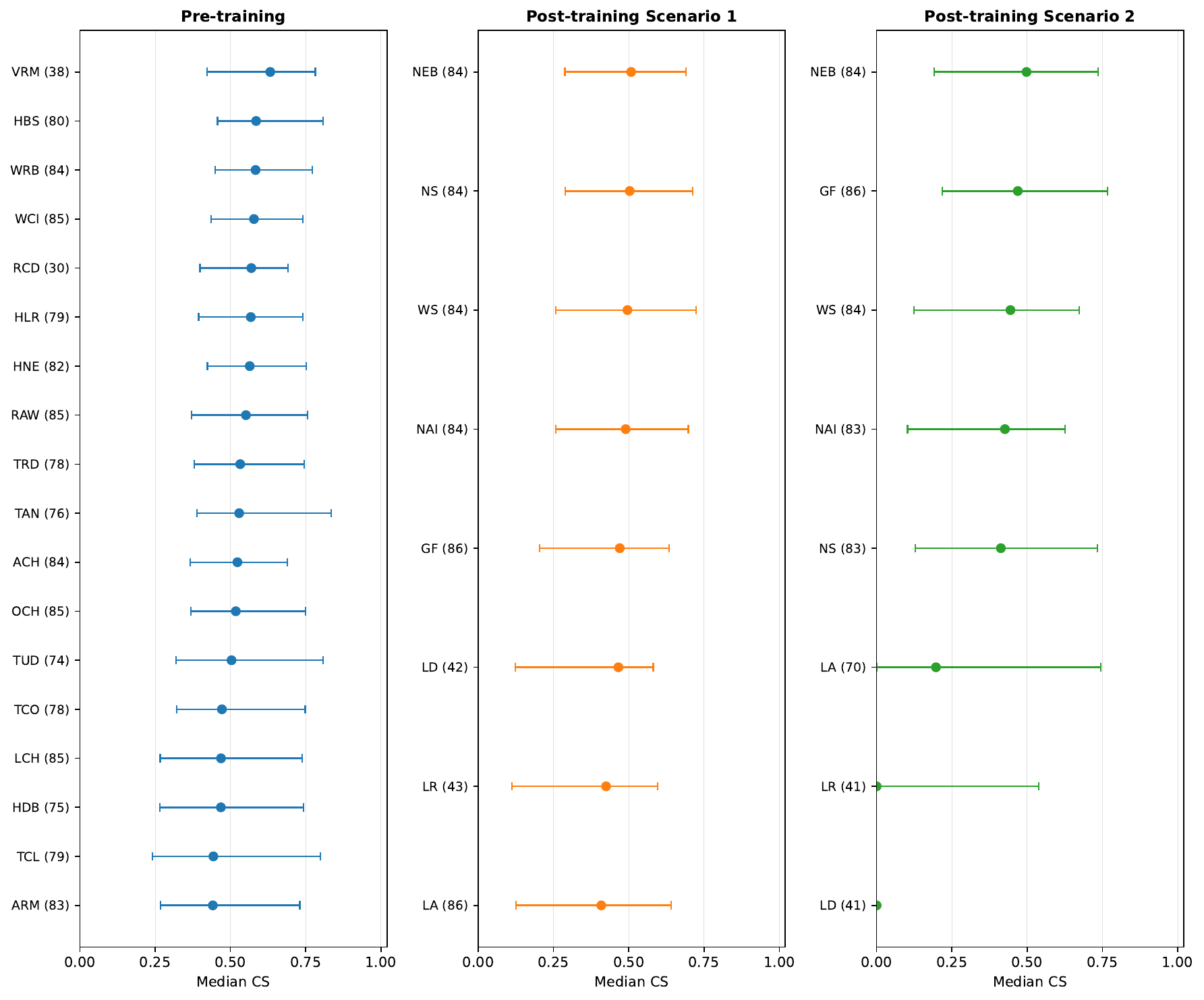}
\caption{Operators ranked by median coupling strength (CS) across bugs. Each dot represents the median of per-bug medians; horizontal bars show the interquartile range (Q1--Q3). The number in parentheses denotes the number of bugs to which each operator was applicable.}
\label{fig:cs_dotplot}
\end{figure*}

\paragraph{Operator-level Analysis:}
To examine whether the approach-level trends hold uniformly across individual operators, we analyse CS at the operator level.
Out of the 24 pre-training operators, four were not applicable to any of the 86 bugs in our benchmark: RCW, RRW, WAB, and OCG.
These operators require specific architectural features (e.g., existing weights regularisation or gradient clipping) that none of the subject models possess.
The remaining 20 operators were applicable to at least one bug.
Among these, two operators were excluded from the ranking due to insufficient coverage (applicability below the 15-bug threshold required for stable quartile estimation): RCP was applicable to only one bug and AAL to seven, which does not allow stable estimation of quartiles.
Figure~\ref{fig:cs_dotplot} ranks the 18 eligible pre-training operators and all eight post-training operators by their median CS across bugs, with interquartile ranges indicating variability.
The number of bugs each operator was applicable to is shown in parentheses.

Among the 18 eligible pre-training operators, applicability ranges from 30 bugs (RCD) to 85 bugs (WCI, RAW, OCH, LCH).
This variation reflects differences in the target artifacts required by each operator: operators targeting universally present components such as the loss function or optimiser apply to nearly all bugs, whereas operators such as RCD require a dropout layer in the original model.
All eight DeepMutation++ operators are applicable to the majority of bugs because they operate directly on trained model parameters, with applicability counts ranging from 41 (LD in Scenario~2) to 86 (GF).

The results reveal that the pre-training advantage observed at the approach level is systematic across individual operators and not driven by a few high-performing operators.
The majority of pre-training operators achieve median CS values above 0.50, with the highest-ranked operators (VRM, HBS, WRB) reaching approximately 0.60.
While the top-ranked weight-level and neuron-level post-training operators (NEB, NS, WS, NAI, GF) achieve median CS values comparable to the lowest-ranked pre-training operators, the distributions differ.
Pre-training operators consistently reach higher third-quartile values, indicating that a larger proportion of bugs achieve strong coupling even under the weakest pre-training operators.
For post-training, NEB ranks highest in both scenarios, reaching a median CS of approximately 0.50.
The weight-level and neuron-level operators (NEB, NS, WS, NAI, GF) consistently outperform the layer-level operators (LR, LA, LD), which occupy the lowest positions in both scenarios.
Scenario~1 produces comparable or slightly higher median CS values than Scenario~2 for most operators, although Scenario~2 exhibits wider interquartile ranges, reflecting greater variability across bugs due to its more aggressive parameterization strategy.

\begin{tcolorbox}[colback=gray!10, colframe=black, boxrule=0.8pt]
\textbf{Finding (RQ\textsubscript{1}):}
Pre-training mutants exhibit significantly higher CS with real faults
than post-training mutants ($p<0.001$, medium to large effect), with some per-bug exceptions.
This advantage is systematic across individual operators.
\end{tcolorbox}

\subsection{RQ\textsubscript{2}: How similar is the detectability of pre- and post-training mutants and real faults?}

\paragraph{Approach-level Analysis:}
To address RQ\textsubscript{2}, we analyze the \emph{behavioral similarity} between mutants and real faults, measured using the \emph{IoU} of their detected test sets. 
IoU captures the degree of overlap in detectability: a higher IoU indicates that the same test inputs tend to expose both a mutant and its corresponding real fault.

\begin{figure}[htbp]
  \centering
  \subfloat[CleanML dataset\label{fig:iou_cleanml}]{%
    \includegraphics[width=1.2\linewidth,height=0.25\textheight,keepaspectratio]{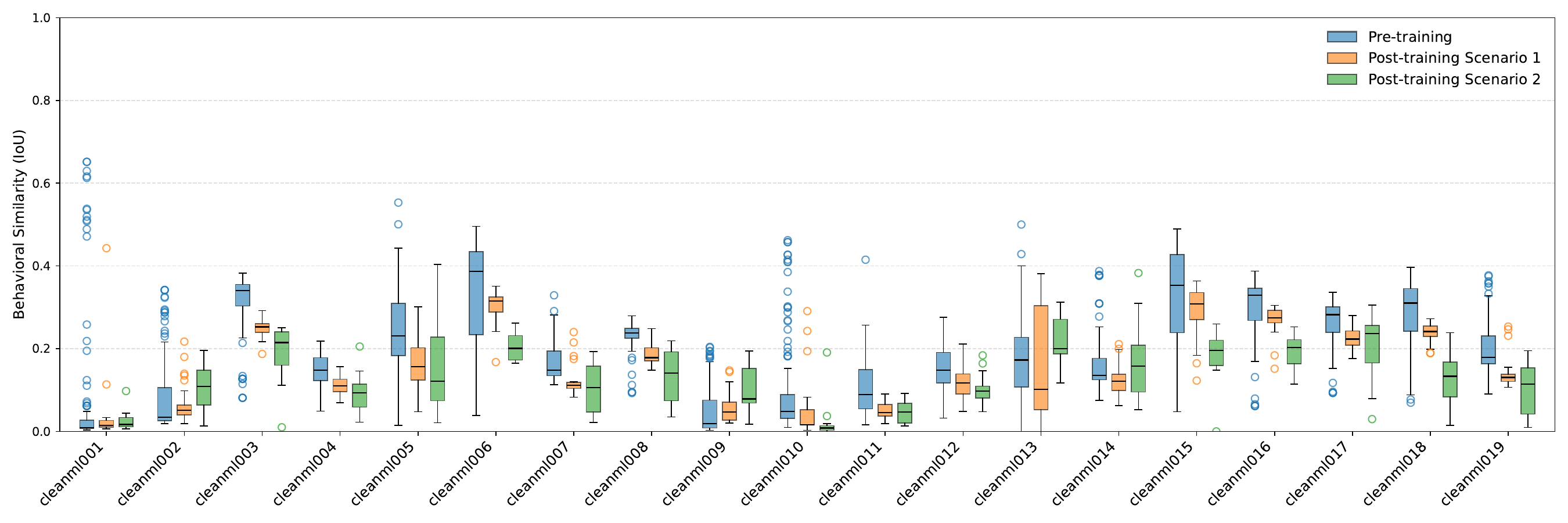}%
  }\\
 
  \subfloat[DeepFD dataset\label{fig:iou_deepfd}]{%
    \includegraphics[width=1.2\linewidth,height=0.25\textheight,keepaspectratio]{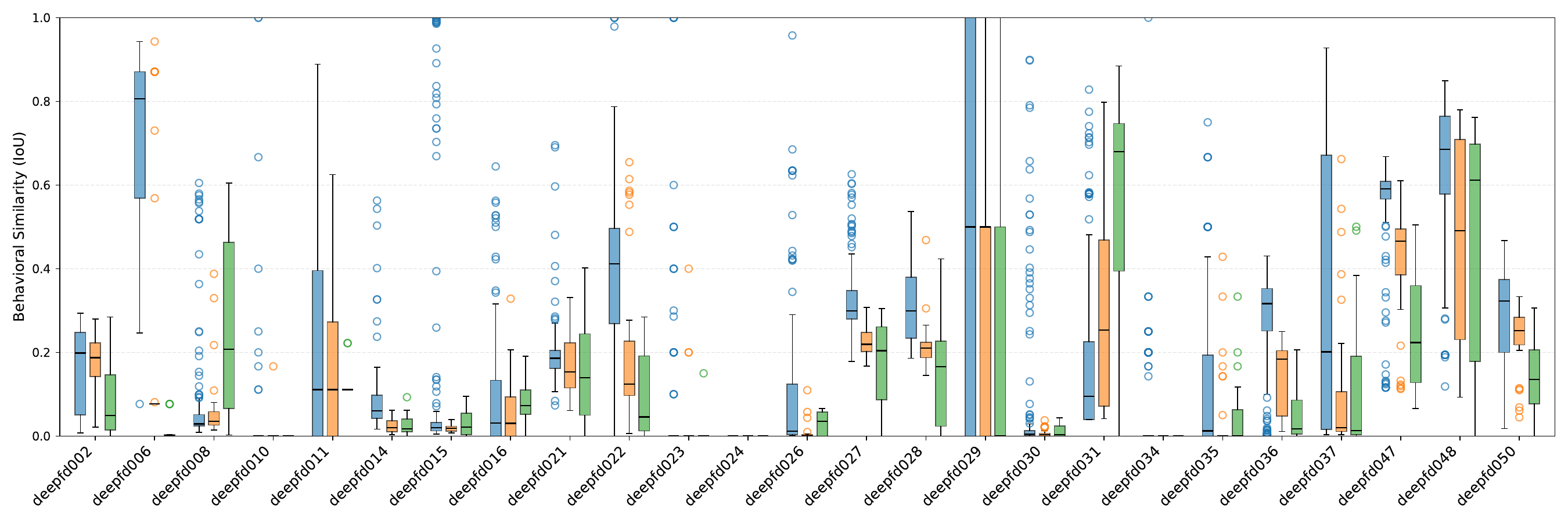}%
  }\\
  \subfloat[DeepLocalize dataset\label{fig:iou_deeplocalize}]{%
    \includegraphics[width=1.2\linewidth,height=0.25\textheight,keepaspectratio]{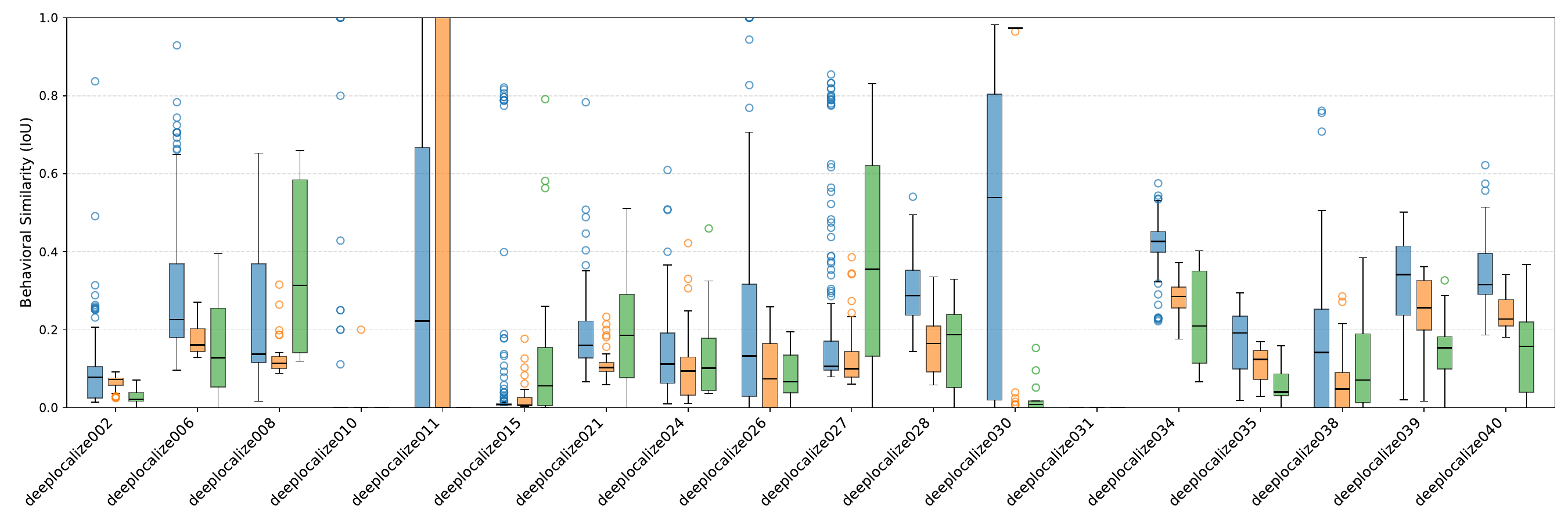}%
  }\\

   \subfloat[defect4ML dataset\label{fig:iou_defect4ml}]{%
    \includegraphics[width=1.2\linewidth,height=0.25\textheight,keepaspectratio]{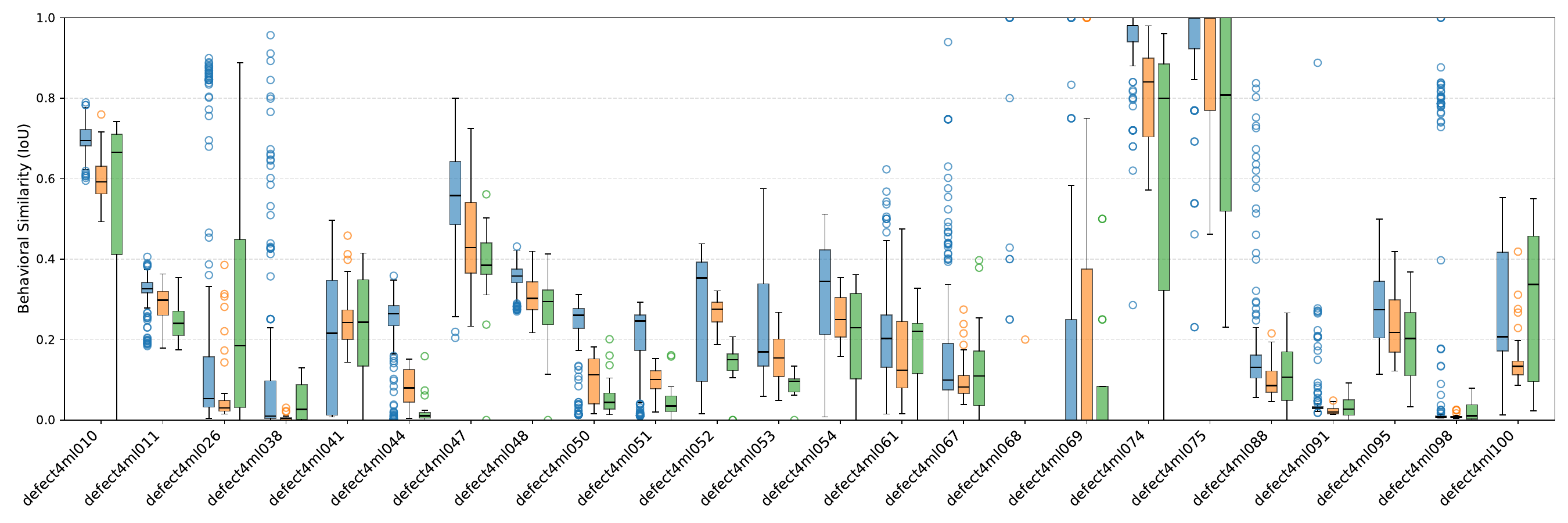}%
  }
  
  \caption{Bug-wise behavioral similarity distributions across datasets. Each bug shows three scenarios: pre-training, post-training scenario 1, and post-training scenario 2. The color coding and legend shown in subfigure~(a) apply consistently to all subfigures~(b)--(d).}
  \label{fig:iou_boxplots}
\end{figure}

Figure~\ref{fig:iou_boxplots} presents bug-wise distributions of IoU values for each dataset: 
CleanML (Fig.~\ref{fig:iou_cleanml}), DeepFD (Fig.~\ref{fig:iou_deepfd}), DeepLocalize (Fig.~\ref{fig:iou_deeplocalize}), and defect4ML (Fig.~\ref{fig:iou_defect4ml}). 
As with CS, each bug is represented by three boxplots corresponding to \emph{pre-training}, \emph{post-training scenario~1}, and \emph{post-training scenario~2} mutants. 

\begin{figure}
    \centering
    \includegraphics[width=1.0\linewidth]{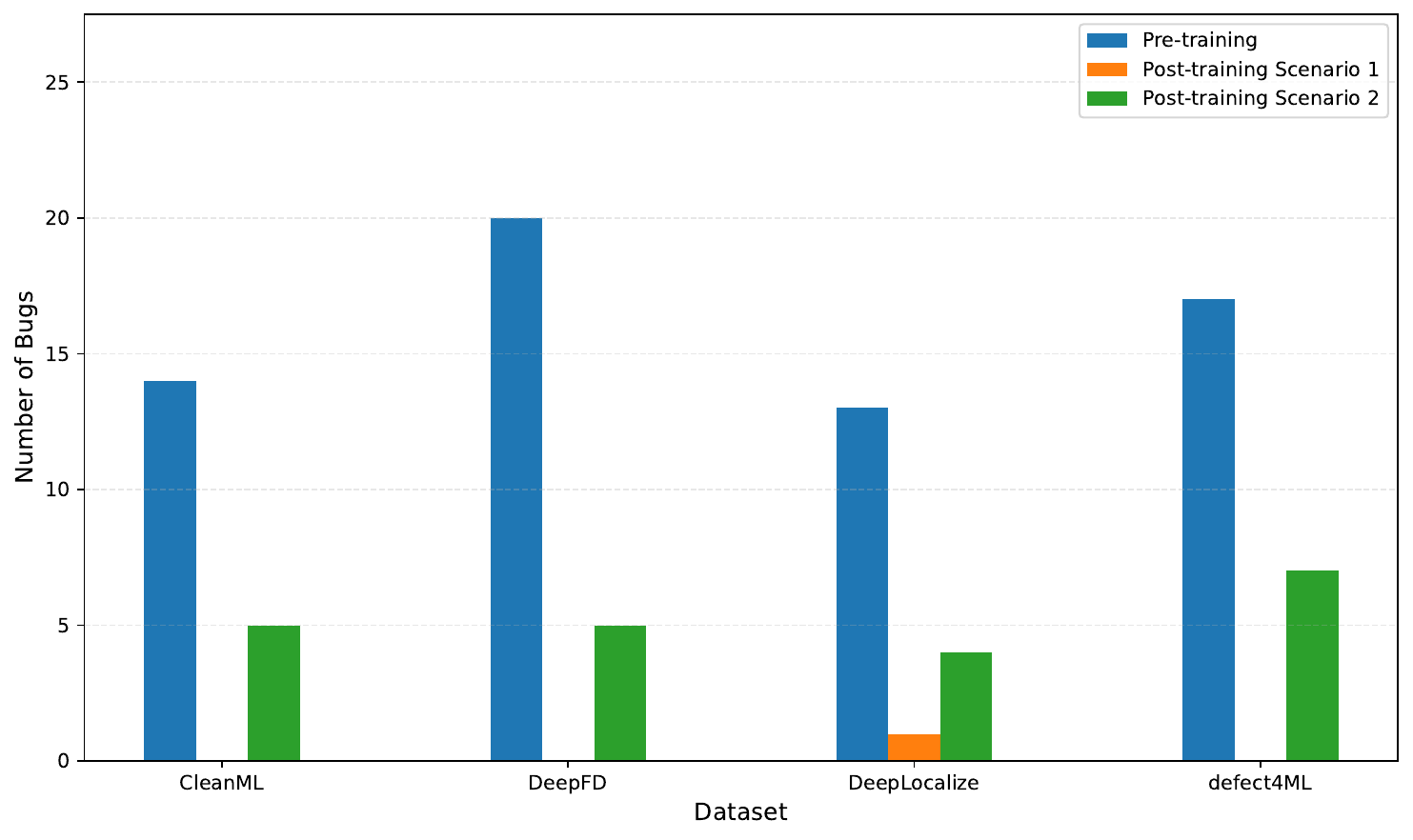}
    \caption{Dataset-level summary of bug-wise cases achieving the highest median behavioral similarity.}
    \label{fig:win_counts_IoU}
\end{figure}

Across datasets, pre-training mutants generally achieve higher IoU values compared to post-training scenarios, indicating stronger similarity in detectability with real faults. The distributions also show that post-training scenario 2 occasionally produces competitive values, but the overall dominance of pre-training is clear. When summarizing results at the dataset level based on which scenario attains the highest median IoU for each bug as illustrated in Fig.~\ref{fig:win_counts_IoU}, ties were resolved in favor of the pre-training scenario, following the same rationale as in RQ\textsubscript{1}. For each bug, the scenario with the highest median IoU was considered the top-performing scenario.

The dataset-level results in Fig.~\ref{fig:win_counts_IoU} highlight the strength of pre-training mutants in terms of detectability overlap: 
\begin{itemize}
    \item \textbf{CleanML (19 data bugs):} Pre-training achieves the highest median IoU for 14 bugs (74\%), post-training scenario 1 for 0 bugs, and post-training scenario 2 for 5 bugs (26\%).
    \item \textbf{DeepFD (25 program bugs):} Pre-training results in the highest median IoU for 20 bugs (80\%), post-training scenario 1 for 0 bugs, and post-training scenario 2 for 5 bugs (20\%). 
    \item \textbf{DeepLocalize (18 program bugs):} Pre-training yields the highest median IoU in 13 bugs (72\%), post-training scenario 1 in 1 bug (6\%), and post-training scenario 2 in 4 bugs (22\%).  
    \item \textbf{defect4ML (24 program bugs):} Pre-training attains the highest median IoU for 17 bugs (71\%), post-training scenario 1 for 0 bugs, and post-training scenario 2 for 7 bugs (29\%).
\end{itemize}

A dataset-level breakdown of IoU confirms that the pre-training advantage holds consistently across all individual datasets.
As with CS, a notable distinction between data and program fault subjects is observed.
CleanML shows lower absolute IoU values across all approaches than the program-fault datasets, as depicted in Fig.~\ref{fig:iou_boxplots}.
The margin of pre-training dominance is also smaller in CleanML, with IoU values consistently below 0.4 and no outliers approaching 1.0.
Program-fault datasets show occasional high-IoU outliers for pre-training mutants, indicating near-complete detectability overlap with real faults in certain cases.

Aggregating across all datasets (86 bugs in total), pre-training mutants achieve the highest median IoU in 64 bugs (74\%), while post-training scenario 1 attains the highest median IoU in 1 bug (1\%) and post-training scenario 2 in 21 bugs (24\%).

Table~\ref{tab:iou_statistical} confirms these observations statistically.
Pre-training mutants achieve significantly higher median IoU than both post-training scenario~1 ($p < 0.001$, $r = 0.845$, large) and post-training scenario~2 ($p < 0.001$, $r = 0.623$, large).
The two post-training scenarios do not differ significantly ($p = 0.069$, $r = 0.241$, small).
Notably, the effect sizes for IoU are larger than those observed for CS (Table~\ref{tab:cs_statistical}), indicating that the gap between pre-training and post-training mutants is even more pronounced in
terms of behavioral similarity than coupling strength.

These results indicate that pre-training mutants are not only more effective as test targets (RQ1), but also more behaviorally similar to real faults in terms of detectability overlap. 
This strengthens their role as realistic substitutes for real DL faults, although the presence of exceptions, particularly in DeepLocalize and defect4ML, shows that post-training mutants can still contribute in certain cases as depicted in Fig.~\ref{fig:iou_boxplots}.

Despite the overall dominance of pre-training, exceptions exist across all datasets. These deviations indicate that the superiority of pre-training mutants is not universal, and we revisit possible explanations in Section~\ref{sec:reasons-for-exceptions}.

\begin{table}[t]
\centering
\caption{Statistical comparison of mutation approaches based on behavioral similarity across all 86 bugs.
Per-bug median IoU values are compared using the Wilcoxon signed-rank test (two-sided).
Effect size is measured by the rank-biserial correlation ($r$).}
\label{tab:iou_statistical}
\begin{tabular}{lccccl}
\toprule
\textbf{Comparison} & \textbf{$n$} & \textbf{$p$-value} & \textbf{$r$} & \textbf{Effect} \\
\midrule
Pre-training vs Post-training Scenario 1              & 86 & $<$0.001 & 0.845 & large \\
Pre-training vs Post-training Scenario 2              & 86 & $<$0.001 & 0.623 & large \\
Post-training Scenario 1 vs Post-training Scenario 2  & 86 &    0.069 & 0.241 & small \\
\bottomrule
\end{tabular}
\end{table}

\begin{figure*}[t]
\centering
\includegraphics[width=\textwidth]{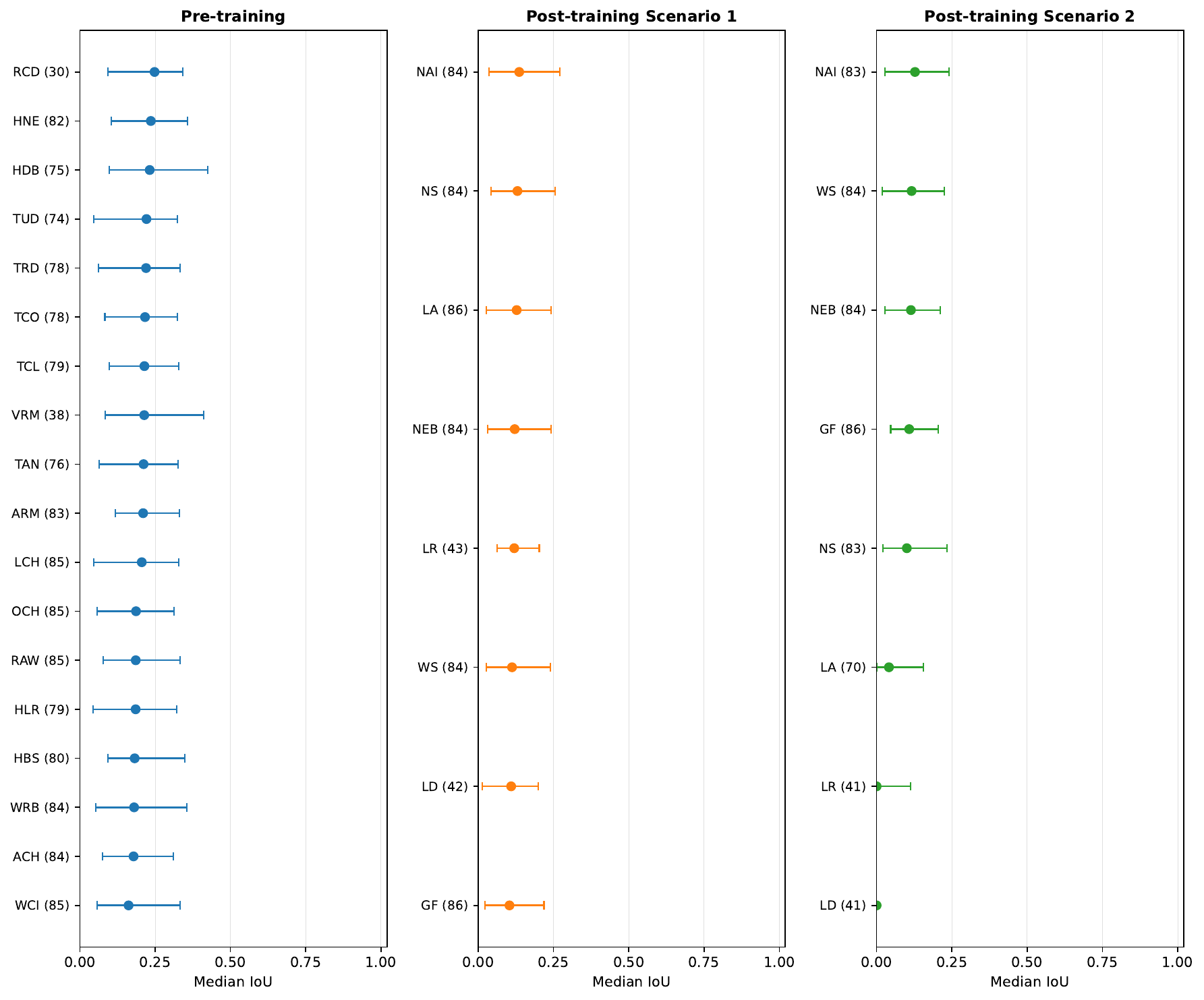}
\caption{Operators ranked by median behavioral similarity (IoU) across bugs. Each dot represents the median of per-bug medians; horizontal bars show the interquartile range (Q1--Q3).  The number in parentheses denotes the number of bugs to which each operator was applicable.}
\label{fig:iou_dotplot}
\end{figure*}

\paragraph{Operator-level Analysis:}
Figure~\ref{fig:iou_dotplot} presents the corresponding operator-level analysis for IoU.
The same operator eligibility criteria as in RQ\textsubscript{1} apply: four pre-training operators (RCW, RRW, WAB, OCG) were not applicable to any bug, and two (RCP, AAL) were excluded from the ranking due to insufficient coverage.

Pre-training operators achieve higher median IoU values than post-training operators across the board.
Pre-training median IoU values range from approximately 0.15 (WCI) to 0.25 (RCD), while post-training values remain below 0.20 in both scenarios.
Among post-training operators, NAI achieves the highest median IoU in both Scenario~1 and Scenario~2.
In Scenario~1, NS and LA follow NAI, whereas in Scenario~2, WS and NEB rank second and third.
The layer-level operators (LR, LA, LD) again occupy the lowest positions, with LD approaching zero in Scenario~2.

As with CS, the top-ranked weight-level and neuron-level post-training operators achieve median IoU values comparable to the lowest-ranked pre-training operators.
However, the pre-training operators exhibit higher third-quartile values even at the bottom of their ranking, indicating stronger tail realism.
Scenario~1 and Scenario~2 produce similar top-end IoU values (NAI reaches approximately 0.17 and 0.18 respectively), but differ substantially at the lower end.
Scenario~2 shows a sharper drop-off for the layer-level operators, with LD and LR falling well below their Scenario~1 counterparts.
This suggests that aggressive parameterisation improves realism for weight-level and neuron-level operators but can degrade it for structural operators.

\begin{tcolorbox}[colback=gray!10, colframe=black, boxrule=0.5pt]
\textbf{Finding (RQ\textsubscript{2}):}
Pre-training mutants exhibit significantly higher behavioral similarity to real faults than post-training mutants ($p<0.001$, large effect), dominating in most bugs.
At the operator level, this dominance holds across all individual pre-training operators.
\end{tcolorbox}

\section{Discussion}
\label{sec:discussion}
The overall message of our results is clear. First, the pre-training mutation approach produces mutants that are significantly more realistic than those generated by the post-training approach. Second, at the per-bug level, CS values tend to exceed IoU values, with CS typically falling within the second to third quartile and IoU within the first to second. These trends are largely consistent across datasets. Nevertheless, exceptions exist: some bugs favor a post-training scenario in terms of realism, and a few edge cases fall into the extreme quartiles for both CS and IoU, or display strong asymmetries such as high CS but low IoU.

In the remainder of this section, we examine these findings in more depth. We first discuss potential reasons for the observed exceptions to the general trends. We then consider the implications of our results for research and practice. Finally, we outline the threats to the validity of our study and their implications for interpreting the results.

\subsection{Reasons for Deviations from the General Trends}
\label{sec:reasons-for-exceptions}

We identify two main types of deviations from the general trends established in Section~\ref{sec:results}: (1) cases where post-training mutants outperform pre-training mutants, and (2) cases where the typical relationship between CS and IoU breaks down.

\subsubsection{Deviations from the Pre-training Superiority Trend}

A small number of bugs appear to favor post-training mutants when judged by per-bug median. Closer inspection shows that most of these are not genuine reversals of the overall trend, but artifacts of summarizing distributions by their centers or of limited oracle sensitivity.

\paragraph{Median superiority that masks tail realism:}
In several cases, post-training edges out pre-training on median CS, yet pre-training still produces highly realistic mutants in the upper tail and often matches or exceeds post-training on IoU:
\begin{itemize}
  \item \emph{Post-training CS median superiority, but both sides reach high tails and pre-training remains competitive on IoU:} deepfd015, deepfd016, deepfd022, deepfd026, deeplocalize026, deeplocalize030, deeplocalize038, defect4ml098. In these bugs, CS for pre-training also reaches values near~1.0, and IoU either favors pre-training by median or shows stronger pre-training outliers.
  \item \emph{Post-training holds a narrow CS-median advantage; pre-training has stronger upper tails and typically leads on the IoU median.} deeplocalize002, deeplocalize006, deeplocalize040, defect4ml053, defect4ml095. Here, the median advantage of post-training is small, while pre-training exhibits many highly coupled or high-overlap outliers.
  \item \emph{Post-training superiority on both medians, but pre-training still yields the most realistic outliers:} deeplocalize021, defect4ml038. Even with dual-median advantages for post-training, the top-performing pre-training mutants exceed all post-training mutants in CS and IoU.
  \item \emph{Post-training shows an advantage only on the IoU median; pre-training leads on the CS median and exhibits stronger IoU tails:} cleanml001, cleanml009, cleanml014, deeplocalize015, defect4ml026, defect4ml041, defect4ml067, defect4ml100.
\end{itemize}
Taken together, these cases show that median comparisons can understate the realism of pre-training: although the medians favor post-training, pre-training mutants routinely achieve CS values near~1.0 and exhibit stronger high-IoU outliers. This indicates competitive or superior realism in the distributional tails.

\paragraph{Oracle-insensitive cases that suppress pre-training’s advantage:}
A few deviations arise when the evaluation oracle provides only a weak discriminatory signal, typically due to high training-test overlap, noisy data, or mild class imbalance that reduces contrast among mutants. Examples include cleanml013 and cleanml018, where severe overfitting leads to saturated accuracies and low-variance losses, limiting the sensitivity of accuracy based detection. When accuracies saturate, the statistical tests used by DeepCrime cannot distinguish behavioral differences, causing many candidate pre-training mutations to be discarded. As a result, only a small subset of potential pre-training mutants is generated, which weakens the basis for coupling analysis. In these cases, apparent post-training advantages do not reflect true superiority but arise because oracle insensitivity prevents the full set of pre-training mutants from being produced.

\paragraph{Summary:}
Most deviations interpreted as favoring post-training are either median-level artifacts that overlook stronger pre-training tails, or occur under weakly discriminative oracles. These findings reinforce pre-training as the stronger default for mutation analysis and highlight the need to evaluate realism using both central and tail behavior, while screening models and datasets for sufficient behavioral observability before mutation analysis.

\subsubsection{Deviations from the CS-IoU Relationship Trend}

The second type of deviation concerns distortions in the expected relationship between CS and IoU. 
Typically, CS moderately exceeds IoU, reflecting that mutants capture a subset of real fault behaviors. 
However, in several edge cases this proportionality breaks down, revealing boundary effects of the oracle and the limits of single-dimensional detection.

\paragraph{High CS but low IoU:}
Bugs such as deepfd015, defect4ml067, and defect4ml098 show strong statistical coupling yet weak detectability overlap, particularly in both post-training scenarios. 
Here, mutants are well correlated with the real fault at a coarse level, showing high CS, but they share only a small portion of failing test inputs, reflected by low IoU. 
This pattern indicates partial behavioral mimicry: the tests that kill these mutants are indeed relevant to the real fault, but many additional inputs that reveal the real fault do not fail on the mutant. 
Such asymmetry highlights the limited coverage of post-training mutants and reinforces the advantage of pre-training, which still produces highly realistic outliers even in these cases.

\paragraph{Both CS and IoU near zero:}
Another group of deviations occurs when both metrics collapse toward zero, reflecting the absence of measurable behavioral variance. 
This is observed in deepfd010, deepfd023, deepfd024, deepfd034, deeplocalize010, deeplocalize031, and defect4ml068. 
In deepfd023, only twenty training samples were available, leading all pre-training mutants to saturate at perfect accuracy (1.0) while the faulty model remained at approximately~0.6. 
In deeplocalize031, extreme class imbalance produced identical accuracy and loss for the faulty, original, and mutant models, leaving no discriminatory signal. 
In deepfd024, both faulty and original models already achieved perfect accuracy, driving all coupling and overlap measures to zero.
The remaining cases, deepfd010, deeplocalize010, and defect4ml068, use the same synthetic dataset with highly separable patterns; despite the faulty version exhibiting a measurable accuracy drop, almost all mutants converged to the same perfect accuracy as the original, yielding zero CS and IoU for all post-training and majority of pre-training mutants. 
Similarly, in deepfd034, differences were observable only in loss values, while accuracy remained fixed at 1.0. 
These systems operate in a saturated condition, where model behavior lacks diversity across mutants. 
In such settings, mutation analysis becomes uninformative unless complemented with multi-dimensional fault-detection criteria and pre-mutation screening for observability.

\paragraph{Both CS and IoU near one:}
At the opposite extreme, defect4ml074 and defect4ml075 achieve CS and IoU medians in the fourth quartile. 
Here, the real fault manifests primarily through differences in loss rather than accuracy. 
Because all models including mutants achieved perfect accuracy, we relied on a loss-based oracle to generate mutants.
These cases demonstrate that even highly deterministic or algorithmically perfect systems can mask subtle behavioral differences unless multi-dimensional metrics are considered.

\paragraph{Summary:}
Taken together, these deviations underline the limitations of single-dimensional fault-detection criteria such as accuracy. 
Across our experiments, accuracy-based oracles frequently failed when models exhibited saturated accuracy (1.0 across all runs), causing mutation tools to terminate prematurely. 
Switching to a loss-based oracle restored observability in several of these cases, whereas in others (e.g., highly trivial or severely imbalanced datasets) even loss showed no variance. 
This diversity of cases indicates that no single metric is sufficient across models. 
Rather than selecting a criterion manually for each case, future DL mutation analysis should employ multi-dimensional detection oracles that adapt to the available behavioral variance, combining accuracy, loss, confidence, and per-class indicators to ensure stable fault observability.
Statistical procedures must also remain robust under boundary conditions and numerical instability, as observed in saturated or trivial regimes. 
Incorporating pre-mutation screening and multi-dimensional detection criteria can substantially improve the effectiveness and reliability of mutation testing for DL systems.

\subsection{Realism of Mutants Across Fault Types}
The results reveal a consistent difference in mutant realism between data and program fault subjects.
Data faults in CleanML originate from data quality issues in the training data, including missing values, outliers, duplicates, and inconsistencies, rather than from explicit code errors.
This structural difference limits the extent to which current mutation operators can replicate the fault mechanism.
Pre-training operators targeting training data partially simulate these properties, yet produce lower coupling and behavioral similarity than observed for program faults.
Post-training operators, which inject weight perturbations into already trained models, have no direct connection to the data-level origin of such faults, further limiting their realism in this context.

Neither approach produces mutants that achieve perfect coupling or near-complete behavioral similarity with data faults.
The absence of upper-tail outliers in CleanML across all approaches reflects an expressiveness gap in current mutation operators for data fault simulation.
This finding highlights the need for enhanced data mutation operators that more faithfully simulate realistic data quality issues.

\subsection{Implications for Mutation Analysis of Deep Learning}
Our findings have several implications for both research and practice.

\paragraph{For Researchers:}  
Our findings point to four concrete research priorities for advancing mutation analysis of DL systems.  

(1) \emph{Design more realistic but efficient mutation operators:}  
Pre-training mutants consistently exhibited higher coupling and behavioral similarity in terms of detectability, confirming their superior realism over the post-training approach.  
However, prior studies report that pre-training can be up to 60× or even 150× more computationally expensive than post-training mutation because each mutant requires full model retraining.  
Developing new post-training or hybrid mutation operators that achieve realism comparable to or better than pre-training at lower cost remains an open research challenge.  

(2) \emph{Develop multi-dimensional fault-detection criteria:}  
Our deviation analysis revealed that single-dimensional kill rules such as accuracy often failed to detect behavioral differences, especially in saturated or imbalanced regimes.  
Future research should explore richer detection criteria that integrate loss deltas, confidence, calibration, and per-class metrics, supported by empirical validation of their sensitivity across datasets and architectures.  

(3) \emph{Establish systematic screening procedures before mutant generation:}  
Similar to traditional mutation testing, where all tests must pass on the original program, DL mutation analysis should begin with a screening step to ensure the model under test is not behaviorally resistant to mutations.  
Screening should confirm that the oracle is sensitive (no accuracy saturation, non-zero loss variance) and that the evaluation data are neither trivial nor severely imbalanced.  
Developing objective metrics for this pre-mutation screening represents a necessary step toward effective DL mutation analysis.

\emph{Enhance mutation operators for data faults:}
Current pre-training operators partially replicate data fault mechanisms but fall consistently short of the coupling and behavioral similarity achieved for program faults.
No existing operator, pre-training or post-training, produces mutants that achieve perfect coupling or near-complete detectability overlap with data faults.
Designing mutation operators that more faithfully simulate data quality issues, such as duplicates, missing values, outliers, and inconsistencies, represents an underexplored but necessary direction for advancing mutation realism in data-centric DL applications.

Beyond mutation testing itself, these findings are relevant to researchers who use mutants as substitutes for real faults in other DL engineering tasks, such as robustness evaluation, automated repair, and test data generation.
If the mutants used in such evaluations are insufficiently realistic, the resulting conclusions may be misleading.

\paragraph{For Practitioners:}
Our findings suggest that when evaluating or strengthening DL test inputs, prioritize pre-training mutants: tests that kill them are more likely to expose real faults, and their behavior aligns more closely with real-fault behavior.
Use post-training as a complement rather than a replacement; if using DeepMutation++~\citep{hu_deepmutation_2019}, weight-level and neuron-level operators under the aggressive configuration (Scenario~2; see Sec.~\ref{sec:experimental-setup}) offer the best realism among post-training options, though layer-level operators perform better under the default configuration (Scenario~1).

Within each approach, the operator-level results further refine this guidance.
Among pre-training operators, the majority achieve median CS above 0.50, with hyperparameter operators (HBS, HLR, HNE) and weight/bias operators (WRB, WCI) consistently ranking among the highest on CS, while training data operators (RCD, HNE, HDB, TUD) rank highest on IoU.
For post-training, weight-level and neuron-level operators (NEB, NAI, WS, GF, NS) substantially outperform layer-level operators (LR, LA, LD) on both metrics.
However, operator rankings differ substantially between the two metrics: for instance, VRM and NEB rank among the highest on CS but fall to mid-range on IoU, whereas RCD and NAI show the opposite pattern.
This divergence reinforces the complementary nature of CS and IoU and indicates that selecting a diverse set of operators -- covering both high-CS and high-IoU operators -- provides the most comprehensive fault coverage.
For teams constrained by training cost, a practical workflow is to generate a limited number of pre-training mutants using the highest-ranked operators and augment them with weight-level and neuron-level Scenario~2 post-training operators to balance realism and efficiency.

\subsection{Threats to Validity}
\label{sec:threats-to-validity}

Like other empirical studies, our work is subject to several threats to validity that may influence the interpretation of results.

\paragraph{Construct Validity:}
Mutant realism was quantified using CS and IoU, which capture complementary dimensions of behavioral similarity.
Both metrics have sound mathematical foundations but may not reflect every aspect of realism, such as structural divergence between models.
Possible bias due to unequal mutant counts across approaches was mitigated by analyzing quartiles and outlier distributions in addition to medians.

\paragraph{Internal Validity:}
Non-determinism in DL training (e.g., random initialization, data shuffling, GPU computation) may have affected observed CS and IoU values.
We mitigated this by fixing random seeds, using identical data splits and environments, and retraining each model five times to capture stochastic variation.
Models producing invalid outputs (e.g., NaN) were excluded, and when single-metric oracles such as accuracy failed due to zero variance, we switched to loss-based evaluation to maintain behavioral observability.

\paragraph{External Validity:}
The study focused on classification faults implemented in Keras/TensorFlow due to tool compatibility and dataset availability.
This constraint limits generalization to other tasks and frameworks such as regression or PyTorch-based systems.
Another threat to generalizability concerns the size of our fault sample. We include 86 reproducible faults. While this does not cover the full spectrum of real-world DL defects, it reflects the practical reality that reproducible faults are rare in DL systems. The findings should therefore be interpreted as representative of the subset of DL faults that can be reliably reproduced and executed in a controlled environment.
Nevertheless, the selection covers four established datasets: CleanML, DeepFD, DeepLocalize, and defect4ML, providing broad coverage across data, architectural, and training-related bug types.

\paragraph{Conclusion Validity:}
Aggregate comparisons between mutation approaches are supported by Wilcoxon signed-rank tests with rank-biserial effect sizes.
At the per-bug level, the analysis additionally relies on medians, quartiles, and comparative counts.
Using medians and quartiles helped to maintain fairness between scenarios with different numbers of mutants, since these measures are less affected by sample size differences.
However, these summaries can understate the role of upper-tail outliers, which in our context represent highly realistic mutants with strong coupling or behavioral similarity.
We explicitly accounted for these cases through a detailed deviation analysis and a tie-breaking strategy that retains meaningful outliers in interpretation.
At the operator level, applicability varies from 1 to 85 bugs across pre-training operators, meaning that operators with lower applicability counts yield less stable quartile estimates.
We mitigated this by excluding operators applicable to fewer than 15 bugs from the operator-level ranking.

\section{Conclusion and Future Work}
\label{sec:conclusion}

This study presented a systematic investigation of mutant realism in DL systems by quantifying coupling and behavioral similarity between mutants and real faults.
We introduced a methodology based on coupling strength and behavioral similarity, capturing complementary aspects of realism under statistical testing conditions.
As part of this work, we analyzed four established real faults datasets (CleanML, DeepFD, DeepLocalize, and defect4ML) and prepared a curated collection of reproducible bugs in a unified environment.
Each selected bug includes both training and evaluation data, enabling consistent and reproducible experimentation across multiple mutation approaches.

The results show that pre-training mutants are significantly more realistic than post-training mutants, achieving higher coupling and greater behavioral alignment with real faults across datasets.
This advantage holds systematically across individual mutation operators, with the majority of pre-training operators outperforming all post-training operators on coupling strength.
Among post-training operators, weight-level and neuron-level operators consistently outperform layer-level operators on both metrics.
However, both our empirical observations and prior studies indicate that generating pre-training mutants is computationally demanding, thereby making large-scale application impractical.
This trade-off highlights the need for new mutation operators that preserve the behavioral realism of pre-training while achieving the efficiency of post-training approaches.

Future work will focus on filling the research gaps identified in this study: designing efficient mutation operators while achieving strong realism, developing multi-dimensional fault-detection criteria that extend beyond accuracy, and defining systematic model-screening procedures to ensure observability before mutation.
Extending the realism assessment to other DL paradigms and frameworks, such as regression and PyTorch-based systems, will further strengthen the generality of these findings.
A further direction is to examine why certain operators achieve higher realism than others.
In particular, it remains open whether high-realism operators succeed because they structurally resemble the corresponding real faults or because they produce behaviorally similar effects through different mechanisms.

\section*{Acknowledgments}
The authors acknowledge the use of the AI-based tool Claude (Anthropic) for assistance with implementation tasks and data curation. All scientific contributions, experimental design, analytical decisions, interpretations, and scholarly content remain entirely the work of the authors.

\section*{Declarations}

\subsection*{Funding}
This work received no specific grant from any funding agency in the public, commercial, or not-for-profit sectors.

\subsection*{Ethical Approval}
Not applicable.

\subsection*{Informed Consent}
Not applicable.

\subsection*{Author Contributions}

Zaheed Ahmed: Conceptualization, Methodology, Software, Data curation,
Investigation, Formal analysis, Visualization, Writing--Original Draft. \\
Philip Makedonski: Validation, Writing--Review \& Editing. \\
Jens Grabowski: Supervision, Visualization, Writing--Review \& Editing. \\
All authors read and approved the final manuscript. \\

\subsection*{Data Availability Statement}
All code, scripts, and processed artifacts used in this study are publicly available at \url{https://github.com/zaheedahmed/dl-mutant-realism}.

\subsection*{Conflict of Interest}
The authors declare that they have no conflict of interest.

\subsection*{Clinical Trial Number}
Not applicable.

%
%

\bibliographystyle{spbasic}      
\bibliography{references}   

%
%

\clearpage
\appendix

\section{Dataset-wise Fault Summary}

\begin{table} [ht]
\centering
\caption{Fault category distribution of the selected defect4ML faults mapped to the~\cite{humbatova_taxonomy_2020} taxonomy.}
\label{tab:defect4ml-fault-categories}
\small
\begin{tabular}{lc}
\toprule
\textbf{Taxonomy Category} & \textbf{\# Faults} \\
\midrule
Model $>$ Model Type \& Properties & 3 \\
Model $>$ Layers $>$ Missing/Redundant/Wrong Layer & 1 \\
Model $>$ Layers $>$ Layer Properties & 1 \\
Model $>$ Layers $>$ Activation Function & 5 \\
\midrule
Training $>$ Hyperparameters & 7 \\
Training $>$ Loss Function & 2 \\
Training $>$ Optimiser & 1 \\
Training $>$ Preprocessing of Training Data $>$ Missing preprocessing & 3 \\
\midrule
Tensors \& Inputs $>$ Wrong Tensor Shape & 1 \\
\midrule
API & 1 \\
\bottomrule
\multicolumn{2}{l}{\footnotesize Composite faults are counted for each distinct defect they contain.} \\
\end{tabular}
\end{table}

\begin{table*}[ht]
\centering
\caption{Descriptive statistics of the selected defect4ML faults.}
\label{tab:defect4ml-details}
\begin{tabular}{llrrl}
\toprule
Fault ID & Training Dataset & Train Size & \# Layers & Fault Type(s) \\
\midrule
defect4ml010 & CIFAR-10 & 45000 & 14 & Wrong network architecture \\
defect4ml011 & CIFAR-10 & 50000 & 13 & Suboptimal number of epochs \\
defect4ml026 & GTSRB & 27839 & 12 & Wrong activation function \\
defect4ml038 & Synthetic dataset & 1890 & 4 & Suboptimal learning rate \\
defect4ml041 & MNIST & 60000 & 8 & Suboptimal network structure \\
defect4ml044 & GTSRB & 27839 & 13 & Epsilon for Adam optimizer \\
defect4ml047 & FER-2013 & 3350 & 29 & Suboptimal batch size \\
defect4ml048 & CIFAR-10 & 50000 & 17 & Wrong filter size for convolution layer \\
defect4ml050 & MNIST & 33600 & 11 & Suboptimal number of epochs \\
defect4ml051 & MNIST & 33600 & 12 & Missing dropout layer \\
defect4ml052 & Fashion-MNIST & 48000 & 7 & Suboptimal learning rate \\
defect4ml053 & MNIST & 60000 & 13 & Suboptimal number of epochs \\
defect4ml054 & MNIST & 60000 & 8 & Suboptimal network structure \\
defect4ml061 & Iris & 75 & 3 & Wrong activation function \\
defect4ml067 & Synthetic dataset & 48000 & 8 & \begin{tabular}[t]{@{}l@{}}Wrong loss function; Wrong activation function\end{tabular} \\
defect4ml068 & Synthetic dataset & 20000 & 9 & Wrong activation function \\
defect4ml069 & Custom dataset & 13 & 7 & Missing preprocessing \\
defect4ml074 & Conway's Game of Life & 15000 & 4 & Wrong loss function \\
defect4ml075 & Conway's Game of Life & 15000 & 4 & Suboptimal learning rate \\
defect4ml088 & Custom dataset & 979 & 16 & Wrong activation function \\
defect4ml091 & Custom dataset & 48000 & 9 & Missing preprocessing \\
defect4ml095 & Sonar & 166 & 3 & Wrong API usage \\
defect4ml098 & MNIST & 60000 & 8 & Wrong tensor shape \\
defect4ml100 & MNIST & 5000 & 4 & Missing preprocessing \\
\bottomrule
\multicolumn{5}{l}{\small Synthetic dataset: Generated programmatically. Custom dataset: Loaded from provided file.}
\end{tabular}
\end{table*}

\begin{table}[ht]
\centering
\caption{Fault category distribution of the selected DeepFD faults mapped to the~\cite{humbatova_taxonomy_2020} taxonomy.}
\label{tab:deepfd-fault-categories}
\small
\begin{tabular}{lc}
\toprule
\textbf{Taxonomy Category} & \textbf{\# Faults} \\
\midrule
Model $>$ Layers $>$ Missing/Redundant/Wrong Layer & 3 \\
Model $>$ Layers $>$ Layer Properties & 2 \\
Model $>$ Layers $>$ Activation Function & 16 \\
\midrule
Training $>$ Hyperparameters & 21 \\
Training $>$ Loss Function & 10 \\
Training $>$ Optimiser & 5 \\
Training $>$ Preprocessing of Training Data $>$ Missing preprocessing & 5 \\
Training $>$ Preprocessing of Training Data $>$ Wrong preprocessing & 1 \\
\bottomrule
\multicolumn{2}{l}{\footnotesize Composite faults are counted for each distinct defect they contain.} \\
\end{tabular}
\end{table}

\begin{table*}[ht]
\centering
\caption{Descriptive statistics of the selected DeepFD faults.}
\label{tab:deepfd-details}
\begin{tabular}{llrrl}
\toprule
Fault ID & Training Dataset & Train Size & \# Layers & Fault Type(s) \\
\midrule
deepfd002 & MNIST & 60000 & 6 & \begin{tabular}[t]{@{}l@{}}Wrong activation function; Suboptimal learning rate;\\ Wrong loss function; Suboptimal number of epochs\end{tabular} \\
deepfd006 & Synthetic dataset & 3200 & 9 & \begin{tabular}[t]{@{}l@{}}Wrong activation function; Wrong loss function;\\ Suboptimal number of epochs\end{tabular} \\
deepfd008 & MNIST & 5000 & 4 & \begin{tabular}[t]{@{}l@{}}Missing preprocessing; Wrong activation function;\\ Wrong optimisation function\end{tabular} \\
deepfd010 & Synthetic dataset & 20000 & 8 & \begin{tabular}[t]{@{}l@{}}Wrong activation function; Suboptimal number of epochs\end{tabular} \\
deepfd011 & Custom dataset & 13 & 4 & \begin{tabular}[t]{@{}l@{}}Missing preprocessing; Suboptimal learning rate\end{tabular} \\
deepfd014 & Adult & 26049 & 3 & \begin{tabular}[t]{@{}l@{}}Missing preprocessing; Suboptimal number of epochs\end{tabular} \\
deepfd015 & MNIST & 60000 & 8 & \begin{tabular}[t]{@{}l@{}}Wrong preprocessing; Wrong activation function ($\times$2);\\ Suboptimal batch size\end{tabular} \\
deepfd016 & Iris & 120 & 3 & \begin{tabular}[t]{@{}l@{}}Missing preprocessing; Wrong activation function ($\times$2);\\ Wrong optimisation function\end{tabular} \\
deepfd021 & Custom dataset & 979 & 16 & \begin{tabular}[t]{@{}l@{}}Wrong activation function; Suboptimal number of epochs;\\ Wrong loss function; Wrong output shape\end{tabular} \\
deepfd022 & Synthetic dataset & 800 & 7 & Wrong activation function \\
deepfd023 & Custom dataset & 16 & 4 & \begin{tabular}[t]{@{}l@{}}Missing preprocessing; Wrong activation function\end{tabular} \\
deepfd024 & Iris & 120 & 3 & Wrong activation function \\
deepfd026 & Custom dataset & 800 & 4 & \begin{tabular}[t]{@{}l@{}}Wrong activation function ($\times$2); Wrong loss function;\\ Wrong optimisation function; Suboptimal number of epochs;\\ Suboptimal batch size\end{tabular} \\
deepfd027 & CIFAR-10 & 50000 & 9 & \begin{tabular}[t]{@{}l@{}}Suboptimal number of epochs; Suboptimal batch size\end{tabular} \\
deepfd028 & Titanic & 712 & 6 & \begin{tabular}[t]{@{}l@{}}Wrong activation function; Wrong loss function\end{tabular} \\
deepfd029 & Custom dataset & 17 & 4 & Wrong loss function \\
deepfd030 & Synthetic dataset & 800 & 7 & \begin{tabular}[t]{@{}l@{}}Wrong activation function; Wrong loss function\end{tabular} \\
deepfd031 & Synthetic dataset & 8000 & 4 & \begin{tabular}[t]{@{}l@{}}Wrong optimisation function; Suboptimal learning rate\end{tabular} \\
deepfd034 & Synthetic dataset & 172 & 11 & Suboptimal learning rate \\
deepfd035 & Iris & 75 & 5 & \begin{tabular}[t]{@{}l@{}}Wrong loss function; Suboptimal number of epochs\end{tabular} \\
deepfd036 & MNIST & 60000 & 12 & Wrong loss function \\
deepfd037 & Higgs Boson & 175000 & 6 & \begin{tabular}[t]{@{}l@{}}Wrong layer; Wrong layer properties;\\ Wrong optimisation function; Suboptimal number of epochs;\\ Suboptimal batch size\end{tabular} \\
deepfd047 & MNIST & 60000 & 5 & Suboptimal number of epochs \\
deepfd048 & Conway's Game of Life & 8000 & 7 & \begin{tabular}[t]{@{}l@{}}Wrong loss function; Suboptimal learning rate;\\ Suboptimal number of epochs\end{tabular} \\
deepfd050 & IMDB & 15000 & 9 & \begin{tabular}[t]{@{}l@{}}Missing dropout layer; Missing layer;\\ Suboptimal number of epochs\end{tabular} \\
\bottomrule
\end{tabular}
\end{table*}

\begin{table}[ht]
\centering
\caption{Fault category distribution of the selected DeepLocalize faults mapped to the~\cite{humbatova_taxonomy_2020} taxonomy.}
\label{tab:deeplocalize-fault-categories}
\small
\begin{tabular}{lc}
\toprule
\textbf{Taxonomy Category} & \textbf{\# Faults} \\
\midrule
Model $>$ Model Type \& Properties & 2 \\
Model $>$ Layers $>$ Missing/Redundant/Wrong Layer & 4 \\
Model $>$ Layers $>$ Layer Properties & 6 \\
Model $>$ Layers $>$ Activation Function & 12 \\
\midrule
Training $>$ Hyperparameters & 6 \\
Training $>$ Loss Function & 8 \\
Training $>$ Preprocessing of Training Data $>$ Missing preprocessing & 4 \\
Training $>$ Preprocessing of Training Data $>$ Wrong preprocessing & 1 \\
Training $>$ Training Data Quality & 1 \\
\bottomrule
\multicolumn{2}{l}{\footnotesize  Composite faults are counted for each distinct defect they contain.} \\
\end{tabular}
\end{table}

\begin{table}[ht]
\centering
\caption{Descriptive statistics of the selected DeepLocalize faults.}
\label{tab:deeplocalize-details}
\begin{tabular}{llrrl}
\toprule
Fault ID & Training Dataset & Train Size & \# Layers & Fault Type(s) \\
\midrule
deeplocalize002 & MNIST & 60000 & 6 & \begin{tabular}[t]{@{}l@{}}Wrong activation function; Wrong loss function\end{tabular} \\
deeplocalize006 & Synthetic dataset & 48000 & 9 & \begin{tabular}[t]{@{}l@{}}Wrong activation function; Wrong loss function\end{tabular} \\
deeplocalize008 & MNIST & 5000 & 4 & \begin{tabular}[t]{@{}l@{}}Missing preprocessing; Wrong activation function\end{tabular} \\
deeplocalize010 & Synthetic dataset & 20000 & 6 & \begin{tabular}[t]{@{}l@{}}Redundant activation layer;\\ Wrong activation function\end{tabular} \\
deeplocalize011 & Custom dataset & 13 & 4 & \begin{tabular}[t]{@{}l@{}}Missing preprocessing; Suboptimal batch size\end{tabular} \\
deeplocalize015 & MNIST & 60000 & 8 & \begin{tabular}[t]{@{}l@{}}Wrong preprocessing; Wrong loss function;\\ Suboptimal number of epochs\end{tabular} \\
deeplocalize021 & Custom dataset & 979 & 16 & \begin{tabular}[t]{@{}l@{}}Wrong output shape; Wrong activation function;\\ Wrong loss function\end{tabular} \\
deeplocalize024 & Iris & 75 & 3 & Wrong activation function \\
deeplocalize026 & Custom dataset & 799 & 4 & \begin{tabular}[t]{@{}l@{}}Missing preprocessing; Wrong activation function;\\ Wrong loss function\end{tabular} \\
deeplocalize027 & CIFAR-10 & 50000 & 9 & \begin{tabular}[t]{@{}l@{}}Missing preprocessing; Wrong filter size ($\times$2);\\ Wrong kernel initializer ($\times$2); Suboptimal batch size\end{tabular} \\
deeplocalize028 & Titanic & 712 & 6 & \begin{tabular}[t]{@{}l@{}}Wrong activation function; Wrong loss function\end{tabular} \\
deeplocalize030 & Synthetic dataset & 720 & 7 & \begin{tabular}[t]{@{}l@{}}Wrong activation function; Wrong loss function\end{tabular} \\
deeplocalize031 & DDOS-ML-Detection & 1944601 & 7 & \begin{tabular}[t]{@{}l@{}}Wrong training data labels; Wrong layer;\\ Suboptimal number of neurons\end{tabular} \\
deeplocalize034 & Cats vs.\ Dogs & 1999 & 10 & \begin{tabular}[t]{@{}l@{}}Suboptimal network structure; Suboptimal batch size\end{tabular} \\
deeplocalize035 & GTSRB & 27839 & 13 & Wrong activation function ($\times$3) \\
deeplocalize038 & Conway's Game of Life & 8000 & 4 & \begin{tabular}[t]{@{}l@{}}Wrong loss function; Suboptimal learning rate\end{tabular} \\
deeplocalize039 & IMDB & 15000 & 6 & \begin{tabular}[t]{@{}l@{}}Missing dropout layer ($\times$2);\\ Suboptimal number of epochs\end{tabular} \\
deeplocalize040 & Custom dataset & 8000 & 13 & Wrong network architecture \\
\bottomrule
\end{tabular}
\end{table}

\begin{table}[ht]
\centering
\caption{Fault category distribution of the selected CleanML faults mapped to the~\cite{humbatova_taxonomy_2020} taxonomy.}
\label{tab:cleanml-fault-categories}
\small
\begin{tabular}{lc}
\toprule
\textbf{Taxonomy Category} & \textbf{\# Faults} \\
\midrule
Training $>$ Training Data Quality & 19 \\
\bottomrule
\end{tabular}
\end{table}

\begin{table}[ht]
\centering
\caption{Descriptive statistics of the selected CleanML faults.}
\label{tab:cleanml-details}
\begin{tabular}{llrrl}
\toprule
Fault ID & Training Dataset & Train Size & \# Layers & Fault Type(s) \\
\midrule
cleanml001 & Citation    & 2998   & 3 & Duplicates \\
cleanml002 & EEG         & 10486  & 3 & Outliers \\
cleanml003 & Marketing   & 6296   & 3 & Missing values \\
cleanml004 & Movie       & 4453   & 3 & Duplicates \\
cleanml005 & Movie       & 6529   & 3 & Inconsistencies \\
cleanml006 & Company     & 10500  & 3 & Inconsistencies \\
cleanml007 & Restaurant  & 7723   & 3 & Duplicates \\
cleanml008 & Restaurant  & 8405   & 3 & Inconsistencies \\
cleanml009 & Sensor      & 10500  & 3 & Outliers \\
cleanml010 & Titanic     & 624    & 3 & Missing values \\
cleanml011 & Credit      & 8387   & 3 & Outliers \\
cleanml012 & Credit      & 10500  & 3 & Missing values \\
cleanml013 & University  & 148    & 3 & Inconsistencies \\
cleanml014 & USCensus    & 10500  & 3 & Missing values \\
cleanml015 & Airbnb      & 7921   & 3 & Duplicates \\
cleanml016 & Airbnb      & 8308   & 3 & Outliers \\
cleanml017 & Airbnb      & 10500  & 3 & Missing values \\
cleanml018 & KDD         & 2497   & 3 & Outliers \\
cleanml019 & KDD         & 10500  & 3 & Missing values \\
\bottomrule
\end{tabular}
\end{table}

\end{document}